\documentclass[preprint2]{aastex63}

\usepackage{natbib}
\usepackage{url}
\usepackage{amsmath}
\usepackage{appendix}
\usepackage{hyperref}

\begin{document}

\title{A Comparison of Star-Forming Clumps and Tidal Tails in Local Mergers and High Redshift Galaxies}

\correspondingauthor{Debra Elmegreen}
\email{elmegreen@vassar.edu}
\author{Debra Meloy Elmegreen}

\affiliation{Department of Physics \& Astronomy, Vassar College, Poughkeepsie, NY 12604, USA}
\author{Bruce G. Elmegreen}
\affiliation{IBM Research Division, T.J. Watson Research Center, 1101 Kitchawan
Road, Yorktown Heights, NY 10598, USA}
\author{Bradley C. Whitmore}
\affiliation{Space Telescope Science Institute, 3700 San Martin Drive, Baltimore, MD 21218, USA}
\author{Rupali Chandar}
\affiliation{Department of Physics \& Astronomy, University of Toledo, Toledo, OH 43606, USA}
\author{Daniela Calzetti}
\affiliation{Dept. of Astronomy, University of Massachusetts at Amherst, Amherst, MA 01003, USA}
\author{Janice C. Lee}
\affiliation{Infrared Processing and Analysis Center (IPAC), California Institute of Technology, Pasadena, CA 91125, USA}
\author{Richard White}
\affiliation{Space Telescope Science Institute, 3700 San Martin Drive, Baltimore, MD 21218, USA}
\author{David Cook}
\affiliation{Infrared Processing and Analysis Center (IPAC), California Institute of Technology, Pasadena, CA 91125, USA}
\author{Leonardo Ubeda}
\affiliation{Space Telescope Science Institute, 3700 San Martin Drive, Baltimore, MD 21218, USA}
\author{Angus Mok}
\affiliation{Department of Physics \& Astronomy, University of Toledo, Toledo, OH 43606, USA}
\author{Sean Linden}
\affiliation{Dept. of Astronomy, University of Massachusetts at Amherst, Amherst, MA 01003, USA}

\begin{abstract}
The Clusters, Clumps, Dust, and Gas in Extreme Star-Forming Galaxies (CCDG) survey with the Hubble Space Telescope  includes multi-wavelength imaging
of 13 galaxies less than 100 Mpc away
spanning a range of morphologies and sizes, from Blue Compact Dwarfs (BCDs) to luminous infrared galaxies (LIRGs), all with star formation rates in
excess of hundreds of solar masses per year.
Images of 7 merging galaxies in the CCDG survey
were artificially redshifted to compare with galaxies at z=0.5, 1, and 2.  Most
redshifted tails have surface brightnesses that would be visible at z=0.5 or 1 but not at z=2 due to cosmological dimming.
Giant star clumps are apparent in these galaxies; the 51 measured have similar
sizes, masses and colors as clumps in observed high-z systems in UDF, GEMS, GOODS, and CANDELS surveys.
These results suggest that some clumpy high-z galaxies without
observable tidal features could be the result of mergers. The local clumps also
have the same star formation rate per unit area and stellar surface density as
clumps observed at intermediate and high redshift, so they provide insight
into the substructure of distant clumps. A total of 1596 star clusters brighter than $M_V = -9$ were identified within the
boundaries of the local clumps. The cluster magnitude distribution function is a power law with
approximately the same slope ($\sim -1$ for a number-log luminosity plot) for all the galaxies both inside and outside the clumps and independent of clump surface brightness.

\end{abstract}

\section{Introduction}
\label{intro}

Star formation in galaxies in the early universe was driven by a combination of cold flows
\citep[see review in][]{sancisi08,sanchez} and mergers \citep[see
review in][]{consel}. Simulations such as EAGLE \citep{schaye15}, IllustrisTNG
\citep{pillepich18}, and SIMBA \citep{dave19} show both processes and reproduce the
observed atomic and molecular gas properties well \citep{dave20}. 

Disk galaxies are increasingly clumpy and irregular at high redshift
\citep[e.g.,][] {cowie,vdb,elm05,elm07a,elm07b,guo18}, and the gas velocity dispersions are several times higher
than in local disk galaxies \citep[e.g.,][]{tacc,wis15,simons,wis19}. 
Star-forming clumps (often referred to in local galaxies as ``complexes'' or ``star-forming complexes;'' ``clumps'' will be used throughout this paper) that are kpc-size can be resolved out to redshifts z=4--5 \citep{{elm05},{elm07a},{elm07b},{elm09a},{guo15},{guo18}}, but their substructure is not resolved. While clumps in local galaxies can be resolved into clusters, clumps in local isolated galaxies are much less massive (by 100$\times$ or more) than those at high redshift (which are typically $10^8-10^9 M_{\odot}$) since they formed under more quiescent conditions \citep{elm09a}. Local dwarf irregular galaxies are low mass analogs of high redshift clumpy galaxies, although evolve more slowly and with lower star formation rates \citep{elm09b}.

Local merging or strongly interacting galaxies, on the other hand, show extreme star formation not seen in isolated galaxies, including
massive star clusters in the Antennae \citep{whit} and other Luminous Infrared
Galaxies (LIRGS; \citealt{lin,adamo20}). \citet{zara} examined over 1000 clumps in 46
nearby interacting galaxies and nearly 700 clumps in 38 non-interacting spirals
to compare their large-scale properties. From multiwavelength spectral energy
distribution fits for the identified clumps, they found that the star formation
rates per unit area are higher and the clumps tend to be younger in the
interacting than in the non-interacting galaxies. \citet{larson}  studied 810
clumps in 48 local LIRGS from narrowband HST imaging and found that
the sizes and star formation rates span a broad range between that seen in local
quiescent galaxies and in z=1--3 galaxies, with the largest (kpc-size) clumps
resembling those in the high z galaxies. Note that \citet{cava} analyzed a clumpy high
redshift galaxy with multiple gravitationally lensed images, and found that clumps
are smaller (by a factor of 2-3 on average) when viewed with higher magnification.
Thus, clumps with sizes from a few hundred pc to $\sim$ a kpc in nearby galaxies
are appropriate for comparison with high z clumps.  If clumps in local mergers have photometric properties similar to those in high z galaxies, then local clumps can provide insight into the substructure of clumps not resolved at high z.

We are also
interested in the appearance of major mergers at high redshift and whether some
distant clumpy galaxies without the usual tidal features can be hidden mergers. 
Mergers might be expected to have distinguishing characteristics such as asymmetry,
tidal tails, multiple nuclei, and complex kinematics. However, at high redshift
these can be difficult to recognize \citep{hopkins10}. For example, simulations by \cite{hung16}
and \cite{simons19} found that line-of-sight velocity distributions inside major
mergers at $z\sim 2$ can resemble those in isolated disks. \cite{blumenthal20} suggested that
visual classifications of mergers have a bias toward massive galaxies in dense
environments where tidal effects are large and surface brightnesses are high; they identified only half of the
mergers in mock catalogs made from the IllustrisTNG-100 simulation.

Nevertheless, several studies have found the merger fraction as a function of
redshift using various combinations of features. \cite{xu20} found, on the basis of
asymmetry, color, and comparisons to the Illustris simulation, that for high mass
galaxies, $>10^{11.3}\;M_\odot$, at redshift $z\sim0.5$, 20\% are star-forming and
of those, 85\% are mergers. \cite{duncan19} measured pair counts at 5 to 30 kpc
separation and found the comoving major merger rate was constant to $z=6$ at masses
greater than $10^{10.3}\;M_\odot$ and that the growth of these galaxies by mergers
was comparable to the growth by star formation at $z>3$.  \cite{oleary20} showed
that major mergers are more frequent at high mass ($10^{11}\;M_\odot$) and low
redshift, where they can dominate the mass growth of these galaxies over {\it in
situ} star formation by up to a factor of $\sim10$. \cite{ventou19} combined
separation and velocity information for mergers in Illustris and compared that to
MUSE data in several deep fields, concluding that 25\% of close pairs at mass
greater than $10^{9.5}\;M_\odot$ are major mergers at redshifts of 2 to 3, with a
decreasing fraction beyond this. \cite{ferreira20} used Bayesian deep learning
models applied to large galaxy samples and simulations to suggest that the fraction
of mergers may be $\sim$3\% at z=0.5, increasing to $\sim$7\% at z=1 and $\sim$22\%
at z=2.

Other methods have been used to identify mergers also. \cite{ribeiro17} looked at $\sim 1200$ galaxies
 at z=2 -- 6 and found that most of the clumpy galaxies
 have only two clumps.
They also noted that these two-clump galaxies have the biggest clumps. They then
derived a major merger fraction of 20\% by considering that the large clumps are
merging galaxies and the small clumps are {\it in situ} star formation. \cite{zanella19}
also considered extended emission clumps or what they refer to as ``blobs'' around clumpy galaxies at z=1--3 and
proposed that peripheral blobs are companions while compact blobs are {\it in situ}
star-formation. The galaxy merger fraction in their sample was less than 30\% for
typical mass ratios of 1:5.

Tidal tails are another indication of mergers. \cite{wen16} studied long tidal
tails in the COSMOS field and, when compared to galaxy pairs, suggested that around
half of disk galaxy mergers at $z<1$ produce detectable tails. An early catalog of
100 galaxies out to $z\sim1.4$ with various types of tidal features was given by
\cite{elm07a}; the disks and tidal tails in that study were about a factor of 2
smaller than local disks and tails. This smaller size is consistent with the
observation of shorter merger times at high redshift
\citep[e.g.,][]{snyder17,mantha18}. \cite{elm06} also cataloged ring galaxies and
curved chain galaxies that could be signs of interactions. These studies indicate
that out to $\sim z=1$, mergers often show observable signs of the interaction. At
higher redshifts, tidal features become difficult to observe.

Another reason major mergers are difficult to recognize at high redshift is that
they rarely trigger a large excess of star formation. Most nearby mergers do not have a
large excess in star formation rate either \citep{knapen15}, which is unlike local
ULIRGS, which are late-stage mergers and do trigger significant star formation
\citep[e.g.,][]{sanders96,tacconi02}. For high redshift, \cite{martin17} used
cosmological simulations to show that even with major and minor mergers, most star
formation in a galaxy is {\it in situ} with only 35\% at $z\sim3$ and 20\% at
$z\sim1$ from enhancements during a merger. Similarly, \cite{mundy17} studied a
sample of mass-selected galaxy pairs and found that star formation during major
mergers represents only 1\% to 10\% of the stellar mass growth at high $z$.
\cite{fensch17} used pc-resolution hydrodynamical simulations to compare star
formation rates in merging gas-rich galaxies.  They showed that high-redshift
mergers have smaller excess star formation rates than low-redshift mergers by about
a factor of 10 because of the generally high velocity dispersion and instability
level in high redshift galaxies even without a merger. \cite{pearson19b} used
convolutional neural networks to identify binary mergers in the SDSS, KiDS, and
CANDELS  surveys. They examined excess star formation rates above the galaxy main
sequence and found only a factor of 1.2 more star formation for merging galaxies.
\cite{wilson19} studied 30 galaxy pairs within 60 kpc and 500 km s$^{-1}$ of each
other at redshifts of 1.5 to 3.5 (12 were major mergers) and found no significant
star formation rate enhancement.

Among galaxies with excessive star formation rates (i.e., significantly
above the star formation main sequence), however,  most are major mergers. \cite{cibinel19}
studied close pairs and morphological mergers on and above the star formation main
sequence up to z=2 and found a merger fraction greater than 70\% above the main
sequence, although as in the other studies, the morphological mergers on the main
sequence did not have much excess star formation.

In this paper, we examine the properties of large-scale star-forming clumps and tidal tails in
7 nearby major mergers in the Hubble Space Telescope CCDG
survey. We measure the clump photometric properties for comparison with previously observed high z clumps.  
We further consider whether high z clumps might
contain massive star clusters that are not directly visible in most surveys, using HST observations of 
the local mergers to identify star clusters inside their clumps. By artificially redshifing the local mergers
to mimic their appearance at z=0.5, 1, and 2, we consider whether the comparatively
poor angular resolution and surface brightness dimming at high redshift make major
mergers resemble isolated clumpy galaxies. 

Section \ref{data} presents
the data used in this study, Section \ref{highz} describes how the images are
artificially redshifted to compare with previous observations of galaxies in
moderate to high redshift samples, Section \ref{complex} presents the photometry
for clumps and compares them with properties of clumps in high
redshift galaxies, and Section \ref{star} considers the star clusters within each clump.
Section \ref{tail} discusses the tidal tail features in local and distant systems,
Section \ref{dustgas} discusses extinction effects and gas properties, and Section
\ref{conc} summarizes our conclusions.

\section{Data}
\label{data} The Hubble Space Telescope survey of Clusters, Clumps, Dust, and Gas
in Extreme Star-Forming Galaxies (CCDG)  by \citet{chandar21} includes 13 extreme
star-forming galaxies in the nearby universe. For this paper, we used
multi-wavelength observations of 7 merging systems in the survey:  Arp 220, ESO
185--IG13, Haro 11 (ESO 350--IG038), NGC 1614, NGC 2623, NGC 3256, and NGC 3690.
Arp 220 is an Ultraluminous Infrared Galaxy (ULIRG), NGC 1614, NGC 2623, NGC 3256,
and NGC 3690 (Arp 299) are LIRGS, and ESO 185--IG13 and Haro 11 are blue compact galaxies.
Their distances range from 38 to 90 Mpc. Photometry was done on images observed
with Hubble Space Telescope ACS or WFC3 using the filters F275W, F336W, F435W,
F438W, F555W, and F814W. (Images were also obtained with filters F110W, F130N,
F160W,  although they are not used in this paper.) The images were drizzled and
aligned, with a pixel scale of 0.0396''. Table \ref{T1} lists the galaxies, the
filters and cameras used in this study,  distances, and linear scales, along with
the number of clumps, and number of clusters within clumps.

\section{High redshift appearance}
\label{highz} Following the procedure of \citet{elm09a}, we have Gauss blurred the
HST images to the spatial resolution a galaxy would have at redshifts z=0.5, 1, and
2. Next the images were re-pixelated to mimic how they would appear at a higher z.
Finally, noise was added to account for cosmological dimming, which reduces the
intensity by  $(1+z)^4$.

The Gauss blurring was done as follows. The spatial size of 1 pixel was determined
for redshifts z=0.5, 1, and 2, which corresponds to 240 pc, 318 pc, and 335 pc,
respectively, using the standard cosmological model \citep[Hubble constant
parameter h=0.72;][]{spergel}.  In order for the Gauss-blurred image to have the
same physical scale as the FWHM of a point source (=2.05 px, based on LEGUS
measurements for HST images), we determined the number of pixels, S,  required for
each galaxy to have the same physical size as a FWHM, F. Then the Gauss blur
factor, G, was G= $\sqrt(S^2-F^2$). The corresponding Gaussian $\sigma$ for the
blur is G/(8 ln 2), used in the IRAF task {\it gauss}.  Next, we re-pixelated the
images with the IRAF task {\it blkavg}, setting the block average to be the ratio
of the number of pixels in the galaxy to match the number of pixels in the FWHM,
divided by the FWHM.  Finally, we measured the sky rms in the original image and
added noise to the blurred, re-pixelated image  to give it the same ratio of peak
intensity to sky rms as in the original image, using the IRAF task {\it mknoise}.
The output of these steps is an image of the galaxy with the same physical
resolution, pixelation, and noise level as the galaxy would appear at each
redshift.

\begin{figure}
\epsscale{1.}
\includegraphics[width=6.in]{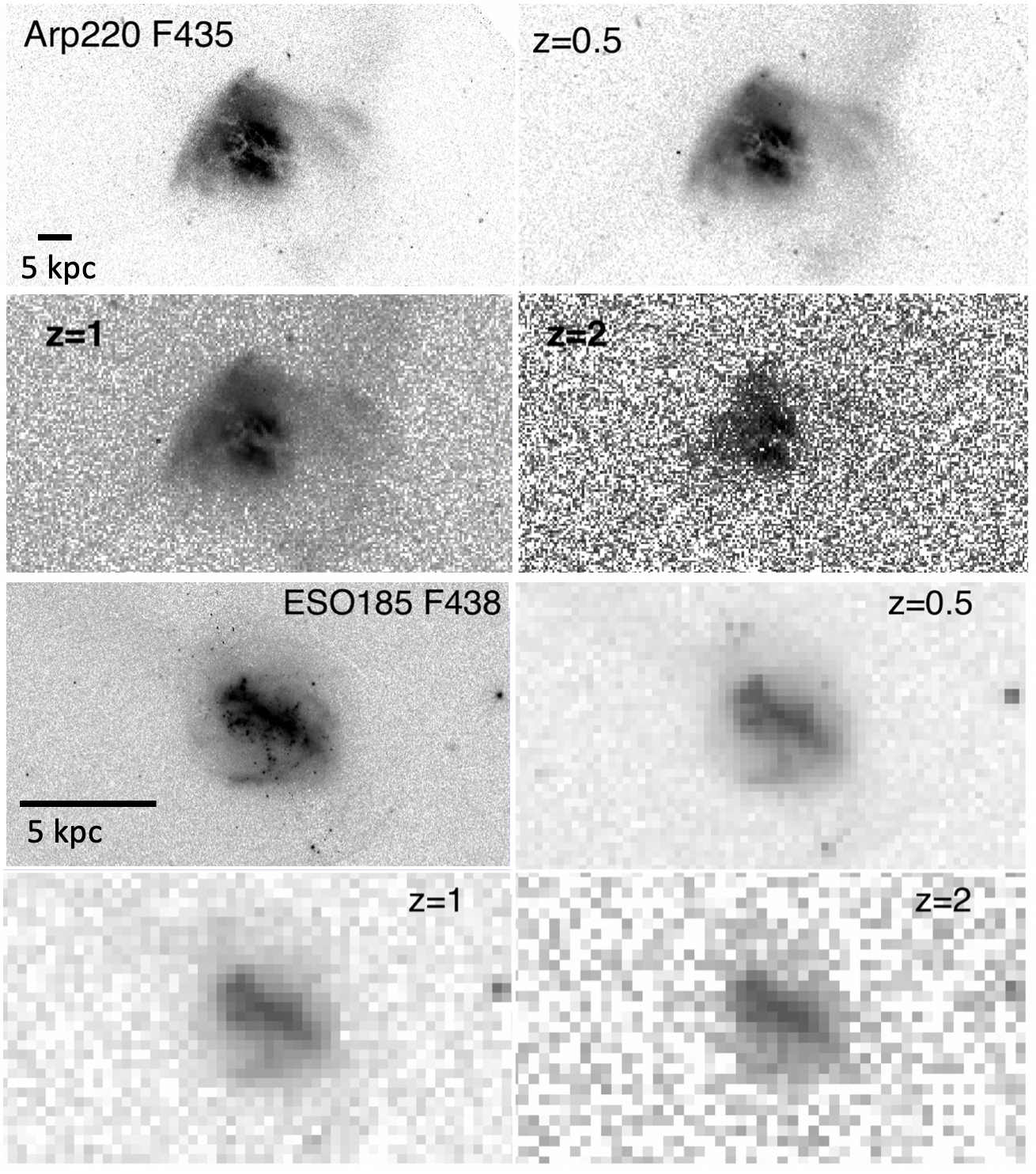}
\caption{Arp 220 and ESO185 are shown in filter F435W, and artificially redshifted to z=0.5, 1, and 2 in the panels, as labeled.  The spatial size of 1 pixel for redshifts z=0.5, 1, and 2 corresponds to 240 pc, 318 pc, and 335 pc, respectively; the pixel scale of the restframe image for each galaxy is in Table \ref{T1}. Black lines indicate a physical scale of 5 kpc. The F435 artificially redshifted image is appropriate for comparison with images in J band, R band, and I band for z=2, 1, and 0.5 galaxies.}
\label{f2}
\end{figure}

\begin{figure}
\epsscale{1.}
\includegraphics[width=6.in]{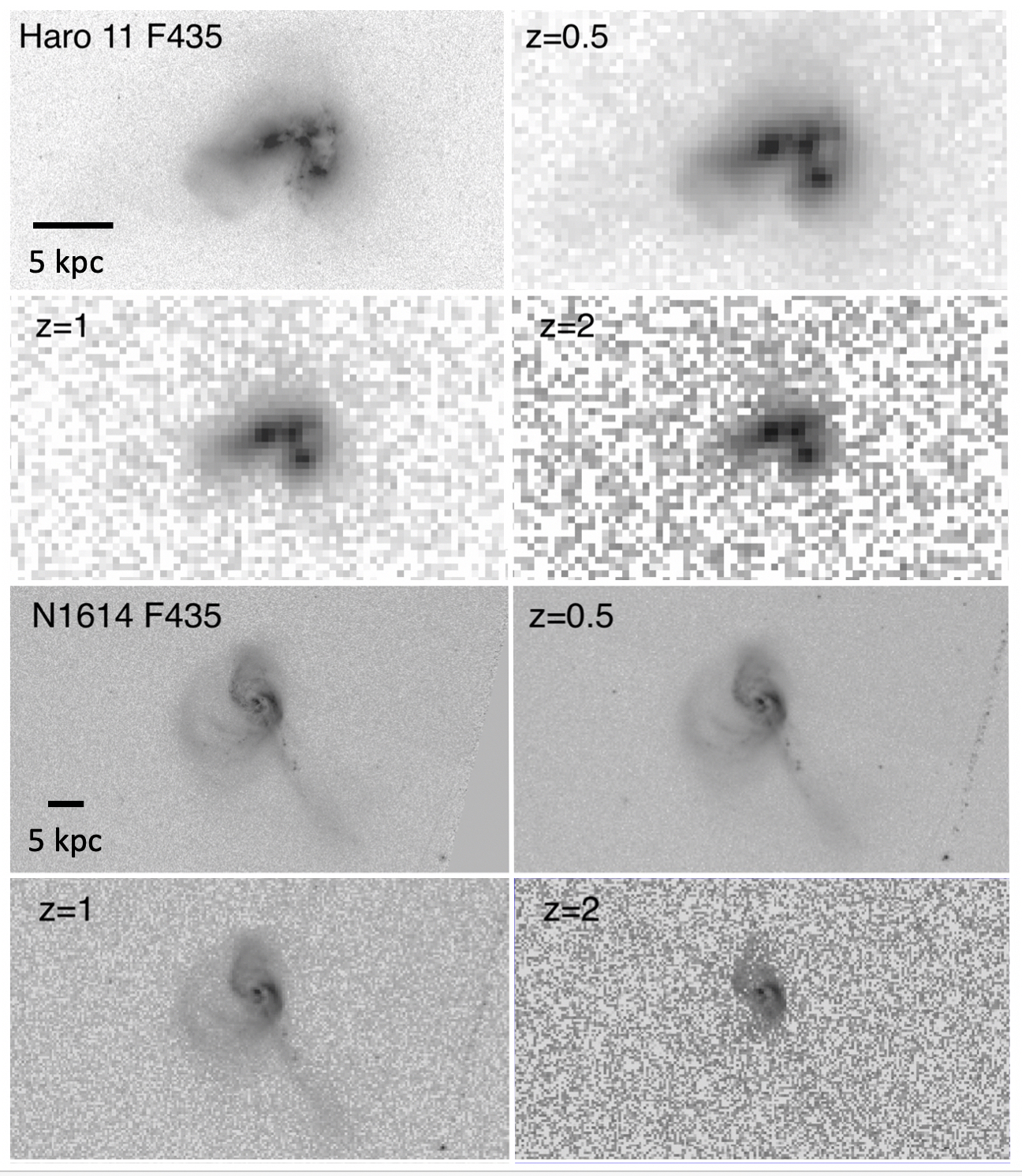}
\caption{Haro 11 and NGC 1614 are shown in filter F435W, and artificially redshifted to z=0.5, 1, and 2 in the panels, as labeled. Black lines indicate a physical scale of 5 kpc. }
\label{f3}
\end{figure}

\begin{figure}
\epsscale{1.}
\includegraphics[width=6.in]{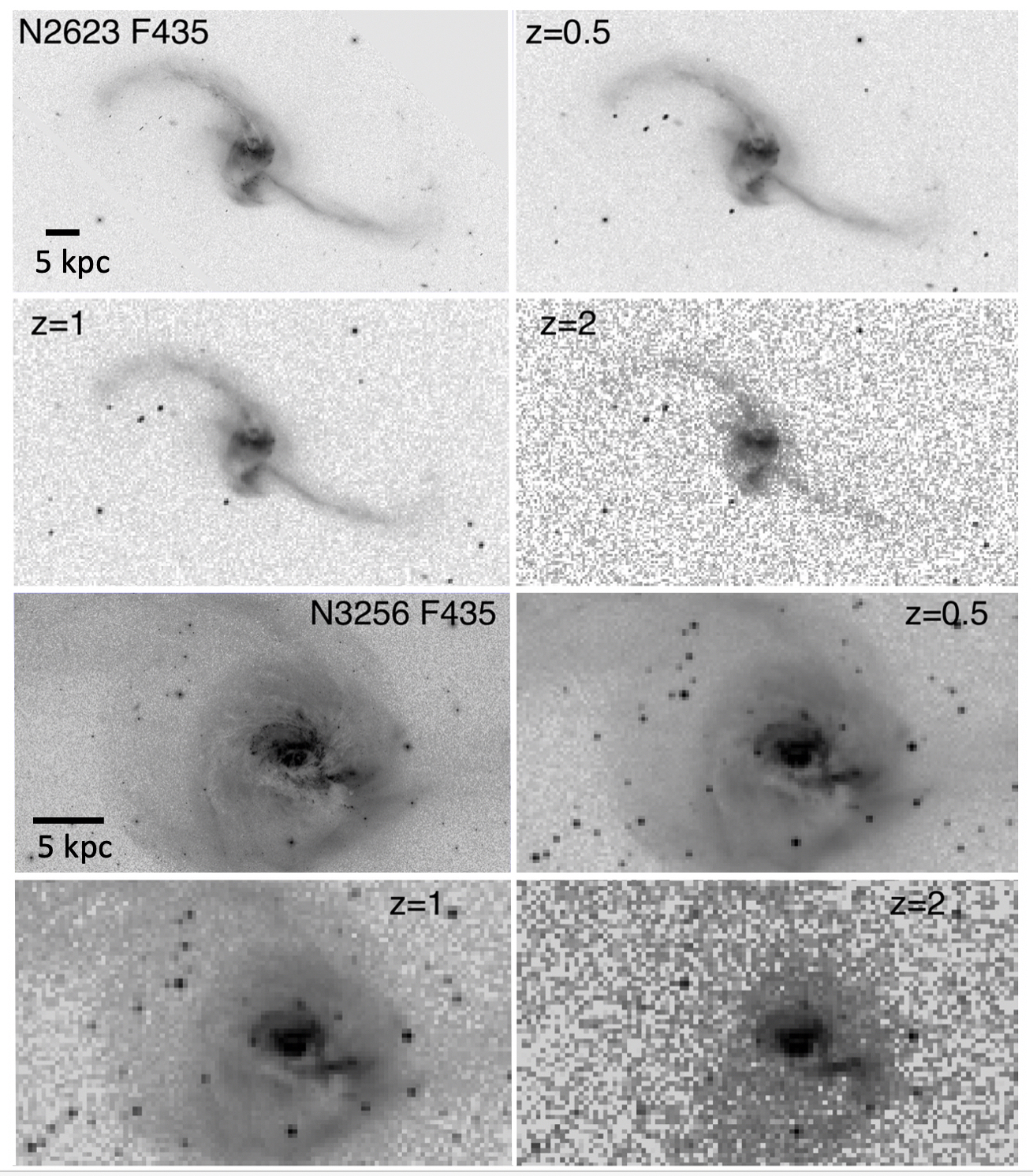}
\caption{NGC2623 and NGC 3256  are shown in filter F435W, and artificially redshifted to z=0.5, 1, and 2 in the panels, as labeled. Black lines indicate a physical scale of 5 kpc. }
\label{f4}
\end{figure}

\begin{figure}
\epsscale{1.}
\includegraphics[width=6.in]{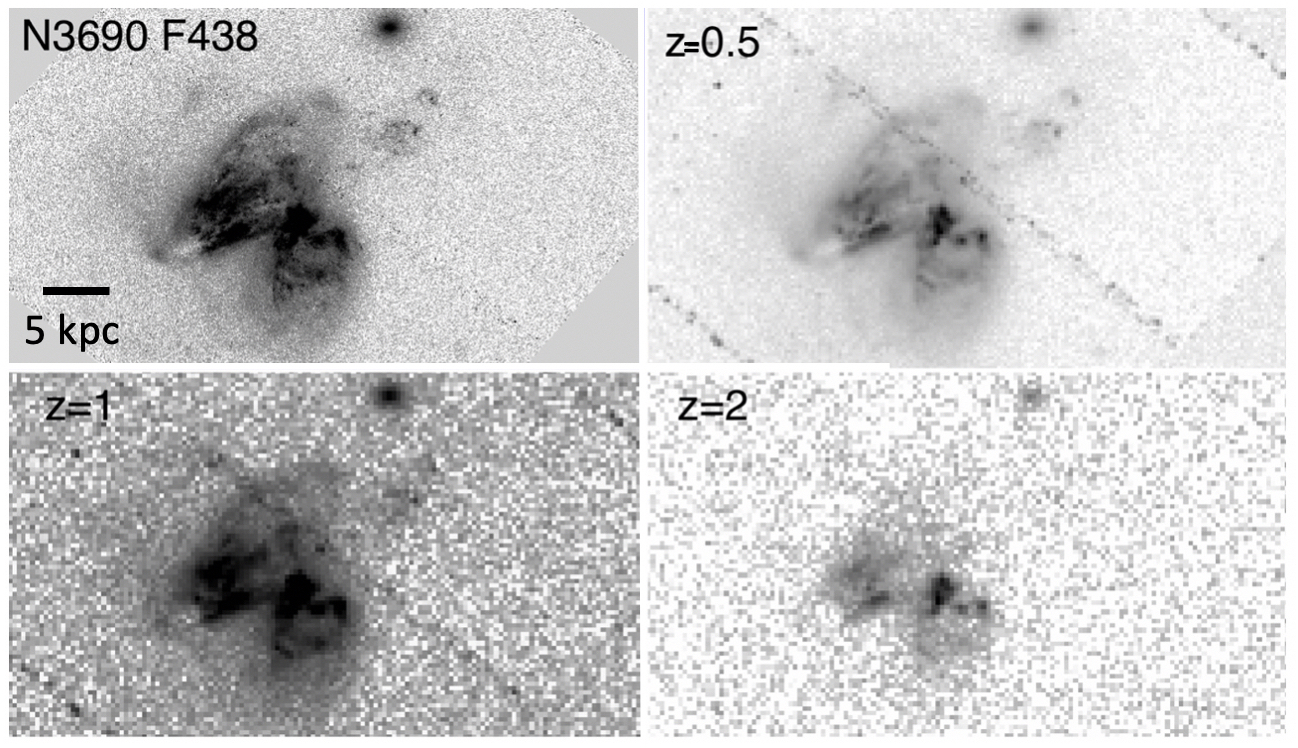}
\caption{NGC 3690 is shown in filter F438W in the upper left panel, and artificially redshifted to z=0.5, 1, and 2 in the panels, as labeled. The black line indicates a physical scale of 5 kpc. }
\label{f5}
\end{figure}

\begin{figure}
\epsscale{1.}
\includegraphics[width=6.in]{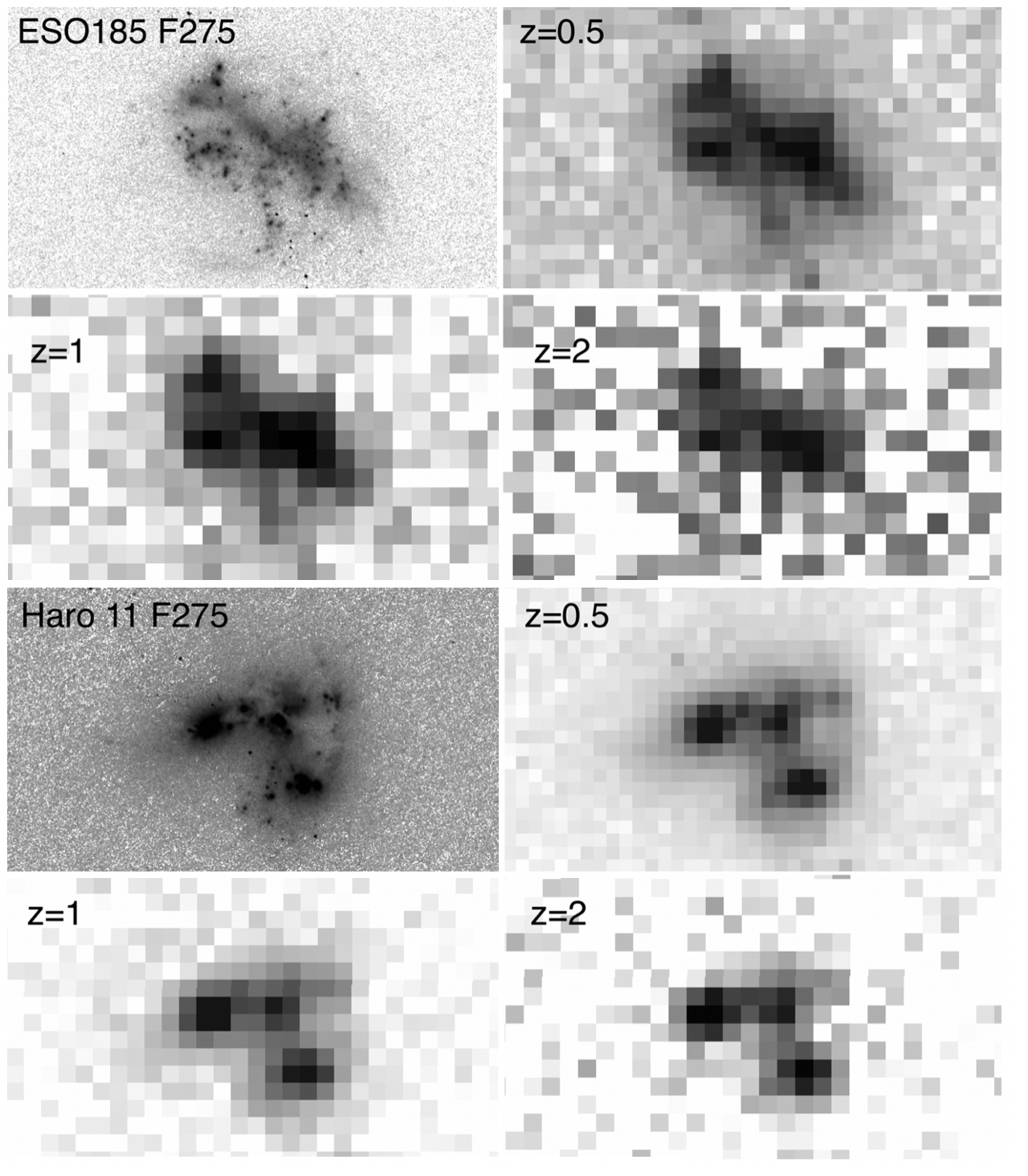}
\caption{ ESO185 and Haro 11 are shown in filter F275W, and artificially redshifted to z=0.5, 1, and 2 in the panels, as labeled. The spatial size of 1 pixel for redshifts z=0.5, 1, and 2 corresponds to 240 pc, 318 pc, and 335 pc, respectively; the pixel scale of the restframe image for each galaxy is in Table \ref{T1}. See Figures 1 -- 2 for 5 kpc scales. The F275 image artificially redshifted is appropriate for comparison with a z=2 galaxy observed in R band, a z=1 galaxy observed in B band, or a z=0.5 galaxy observed in U band.}
\label{f6}
\end{figure}

\begin{figure}
\epsscale{1.}
\includegraphics[width=6.in]{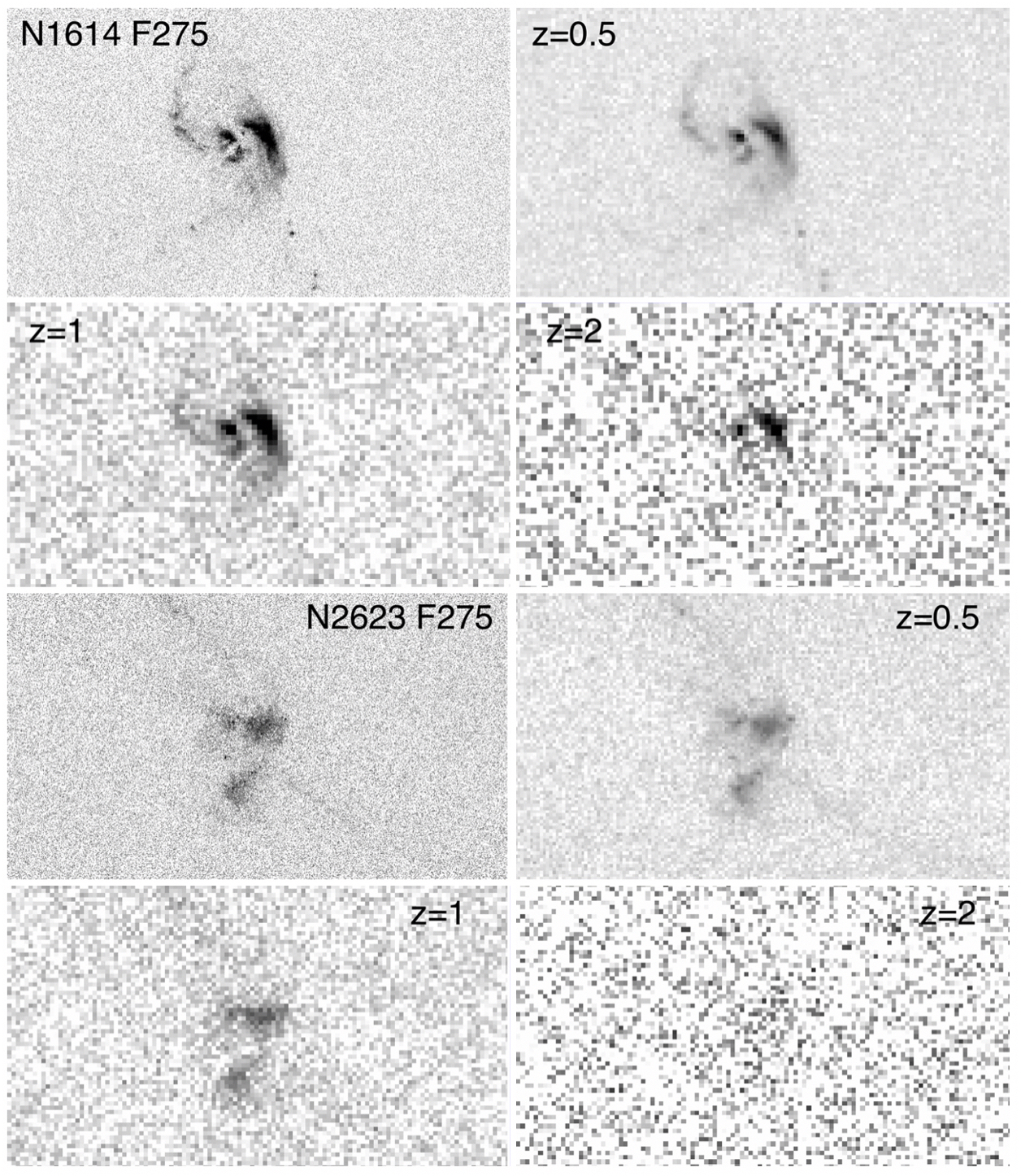}
\caption{NGC 1614 and NGC 2623 are shown in filter F275W, and artificially redshifted to z=0.5, 1, and 2 in the panels, as labeled. See Figures 2 -- 3 for 5 kpc scales.}
\label{f7}
\end{figure}

\begin{figure}
\epsscale{1.}
\includegraphics[width=6.in]{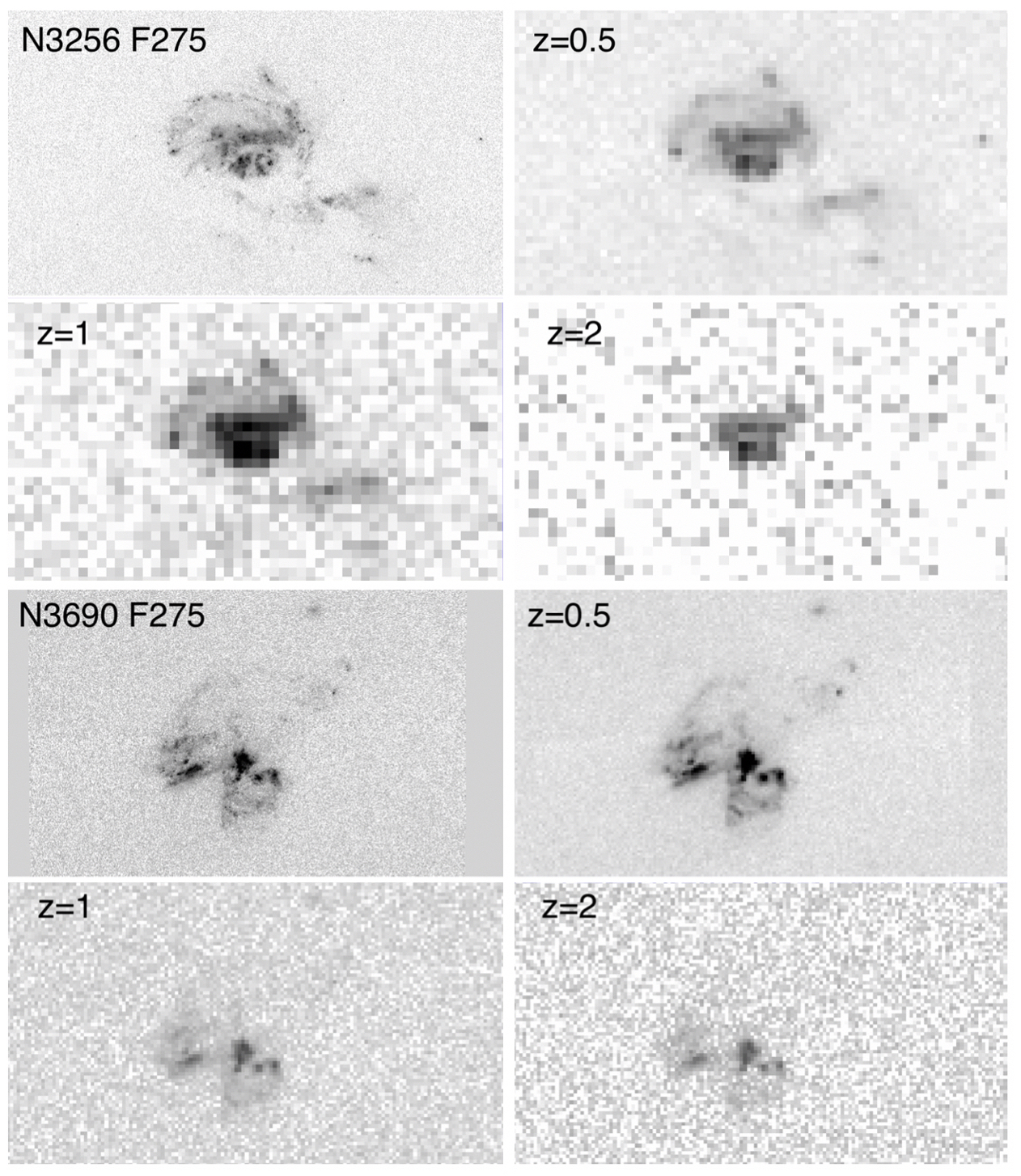}
\caption{NGC 3256 and NGC 3690 are shown in filter F275W, and artificially redshifted to z=0.5, 1, and 2 in the panels, as labeled. See Figures 3 -- 4 for 5 kpc scales.} 
\label{f8}
\end{figure}

The results are shown in Figures \ref{f2} -- \ref{f5} for each galaxy for the F435W
(or 438) images and in Figures \ref{f6} -- \ref{f8} for the F275W images. The F275W
image was very faint for Arp 220 so is not included.

Images in these two filters were selected for ease in comparison with other
observations: the F275W artificially redshifted images are appropriate for
comparison with a z=2 galaxy observed in R band, a z=1 galaxy observed in B band,
or a z=0.5 galaxy observed in U band, while the F435W artificially redshifted
images are appropriate for comparison with J band, I band, and R band images for
galaxies at z=2, 1, and 0.5, respectively.

Merging galaxies out to $z\sim 1$ studied by \citet{elm09a} in the Great
Observatories Origins Deep Survey \citep[GOODS;][]{gia} and Galaxy Evolution from
Morphology and SEDs \citep[GEMS;][]{rix} fields also contained 
clumps. The classified morphologies were ``diffuse,'' which have indistinct
tidal patches or shells of modest size, ``antennae,'' which have two main tidal
arms, ``M51-type,'' which have a long tidal arm pointing to a companion, and
``shrimp-like,'' which have one long and curved clumpy arm connected to a head of
about the same width. The 7 galaxies in the current sample most resemble the
diffuse and antennae morphologies: NGC 2623 is antennae-like, while the other 6 are
diffuse.

Some bright galaxies, like Arp 220, have tidal features that would still be
apparent at higher z, although for most galaxies only the bright centers would
be evident. Haro 11, for example, has the appearance of a clumpy galaxy at higher z;
NGC 2623 would just be a double galaxy at z=2. Several of the galaxies still show
some tidal arms at z=1. The tail surface brightnesses and clump properties will
be compared with high redshift interacting galaxies in more detail below.

\section{Clump Properties}
\label{complex}
\subsection{Clump photometry}
Clumps are identified as extended dense regions. Since clumps are
hierarchical, they subdivide into smaller star-forming regions with higher
resolution. In order to compare clumps in our galaxies with those in high redshift
galaxies, we identified clumps by visual inspection of the z=2 images in the F435W band. 

Measurements were made on the original non-redshifted images since the simulated redshifted images were Gauss-blurred and re-pixelated with artificial noise added, so flux would have been difficult to determine accurately. Instead, contour plots were made on the local images and compared with the z=2 images where the clumps were distinct. In practice, the contours at 10$\times$ sky $\sigma$ on the local images matched the clumps in the high z images, so that is where the boxes were defined for photometry. The IRAF task {\it imstat} was used to determine total
pixels and mean counts per pixel within the boxes. Magnitudes were then determined
for each clump, using the zeropoints from the WFC3 and ACS handbooks. There was
no background subtraction, because the photometric fitting described below included
an underlying component plus the star-forming component, described further below.This is
the same procedure as we used in measuring clumps in high redshift galaxies
\citep{elm09a}, so our results for clumps in our galaxies can be directly
compared to the high redshift results. For the galaxies in this paper, this means
the clump boundaries sometimes contain more than one large clump, since some get blended at high z. 

\begin{figure}
\epsscale{1.}
\includegraphics[width=6.5in]{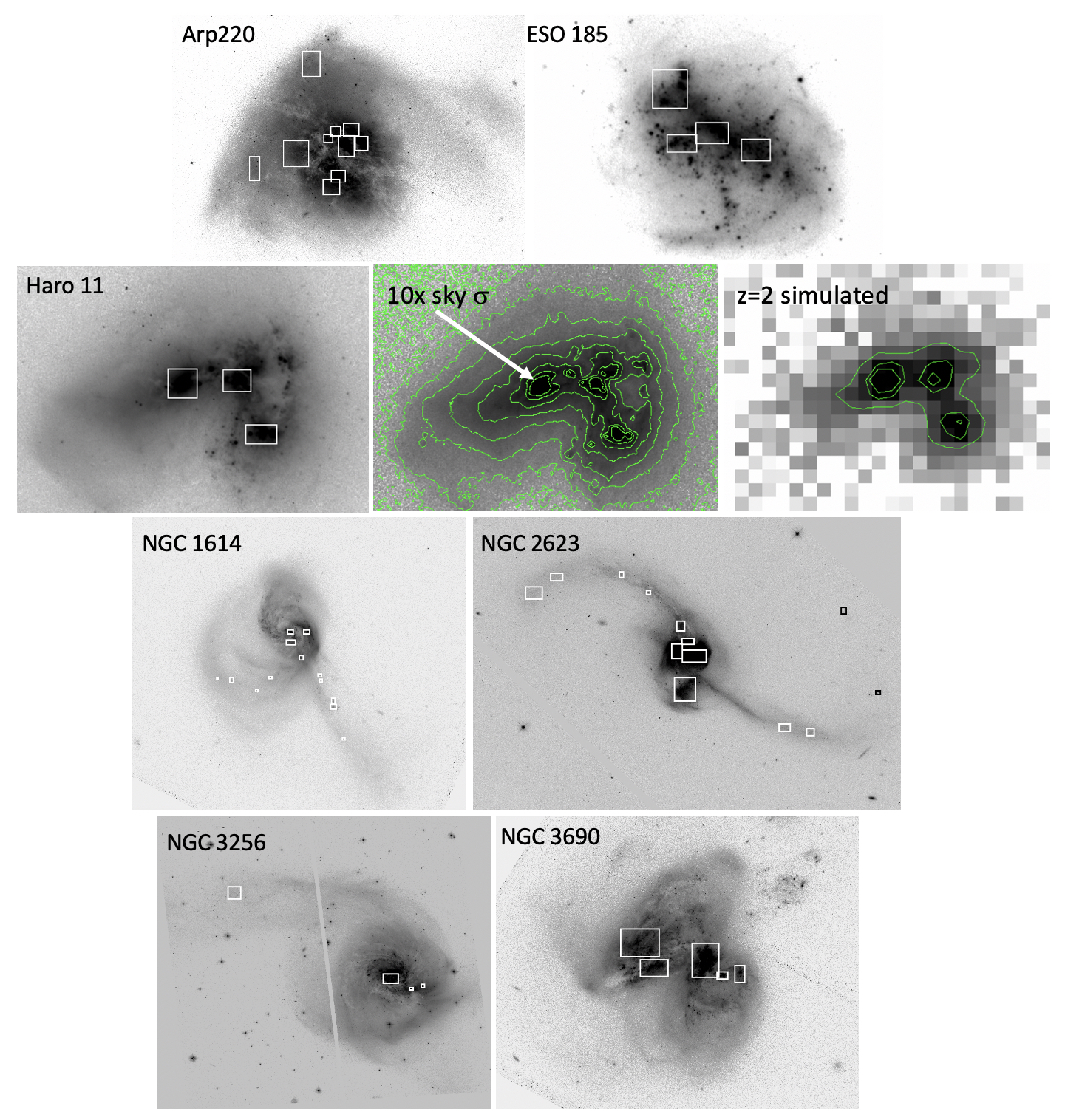}
\caption{The 7 galaxies studied here are shown in the F555W filter. Boxes identify the clumps that were measured; see Figures 1--4 for physical scales. For Haro 11, contour plots are also shown to indicate how clumps were identified. Measurements were done on the non-redshifted images for the boundaries determined from the redshifted images. In practice, these boundaries were at a surface brightness $\sim$ 10x sky $\sigma$ in the F435W image, as indicated in the middle figure. The right-hand contour plot is for the simulated z=2 image F435W image, where the 3 clumps are distinct.  }
\label{fig8galnew}
\end{figure}

 The measured clumps are indicated as boxes in Figure \ref{fig8galnew}. For Haro 11, contour plots are also shown as an example of how clumps were identified. The right-hand contour plot is on the simulated z=2 image F435W image, where the 3 clumps are distinct. The middle contour plot on the original image has an arrow indicating the  $\sim$ 10x sky $\sigma$ contour where the boxes were then drawn.  A total of 51 clumps were identified in the 7 galaxies
(the number in each galaxy is listed in Table \ref{T1}). The average absolute B band magnitude of the clumps for the 7 galaxies in this
sample is -15.4$\pm$2.6, while the average (B-V) color  is $0.4\pm 0.5$. The clump diameters are approximated by the square root of the area of the boxes defined for photometry, and range from 170 pc to 2.9 kpc, with an average of $1.0\pm 0.7$ kpc.

\subsection{Clump ages and masses}
The photometric measurements of the clumps were used to determine ages, masses,
and extinctions through SED fitting, as detailed in \citet{elm09b}. The procedure
multiplied the throughput of each HST filter by an integrated spectrum using
stellar population models from \citet{bruz} with a Chabrier initial mass function
and solar metallicity. The integrated spectrum included two components, a constant
star formation rate for the full span of time in \cite{bruz} representing the
underlying galaxy, plus another constant rate for some variable time representing
burst star formation in the clump.

The ratio of the burst rate to the underlying rate ranged from $10$ to $10^4$ in 12
logarithmic steps, the start time for the burst ranged from $10^7$ to $10^9$ years
in 8 logarithmic steps, and the visual extinction ranged from 0 to 10 mag in
steps of 0.1 mag. For each combination of parameters, the colors of the integrated
spectrum were differenced from the observed clump colors and divided by the
measurement error in that color, and the sum of the squares of these normalized
differences was determined. This sum is considered to be the $\chi^2$ value of the
fit. The best fit parameters were then taken to be the weighted average of the
input parameters with a weighting factor equal to $\exp(-w\chi^2)$, where $w$ is a
parameter chosen to make the weighting factors vary slowly around the maximum
weight.  The measurement error for color was taken to be the square root of the sum
of the squares of the measurement errors for each passband, and the latter were
taken to be the ratio of the standard deviation of the flux count to the flux
itself, averaged for all pixels in the clump (from IRAF {\it imstat}) divided by
the square root of the number of pixels in the clump. The measurement errors are
generally much lower than the color differences among the models, so $w$ was chosen
to be fairly large to keep the exponential weighting factor close to unity; we
choose $w=2^{0.5}\times20$ after some experimentation. The model uncertainty was
determined from the range of model values with the lowest $\chi^2$.

These models do not alone give the clump masses or star formation rates, as they
fit only the star clump colors. The evaluation of mass and absolute magnitude in
each passband comes from a comparison between the observed and model $I$-band flux.
That is, the absolute mass was obtained from the product of the model mass and the
ratio of the observed $I$-band flux to the model $I$-band flux. (The results for
mass in what follows refer to the mass of the young component in the two-component
population fit for a clump, unless otherwise indicated.) Because of a
general ambiguity in stellar population colors between age and extinction, and a
compensation between these ambiguities in the determination of mass, the fitted
masses are considered to be more accurate than the ages and extinctions. The average error in the fits was 0.13 in log mass, 0.33 in log age, and 0.27 in $A_V$.

\begin{figure}
\epsscale{1.}
\includegraphics[width=4.in]{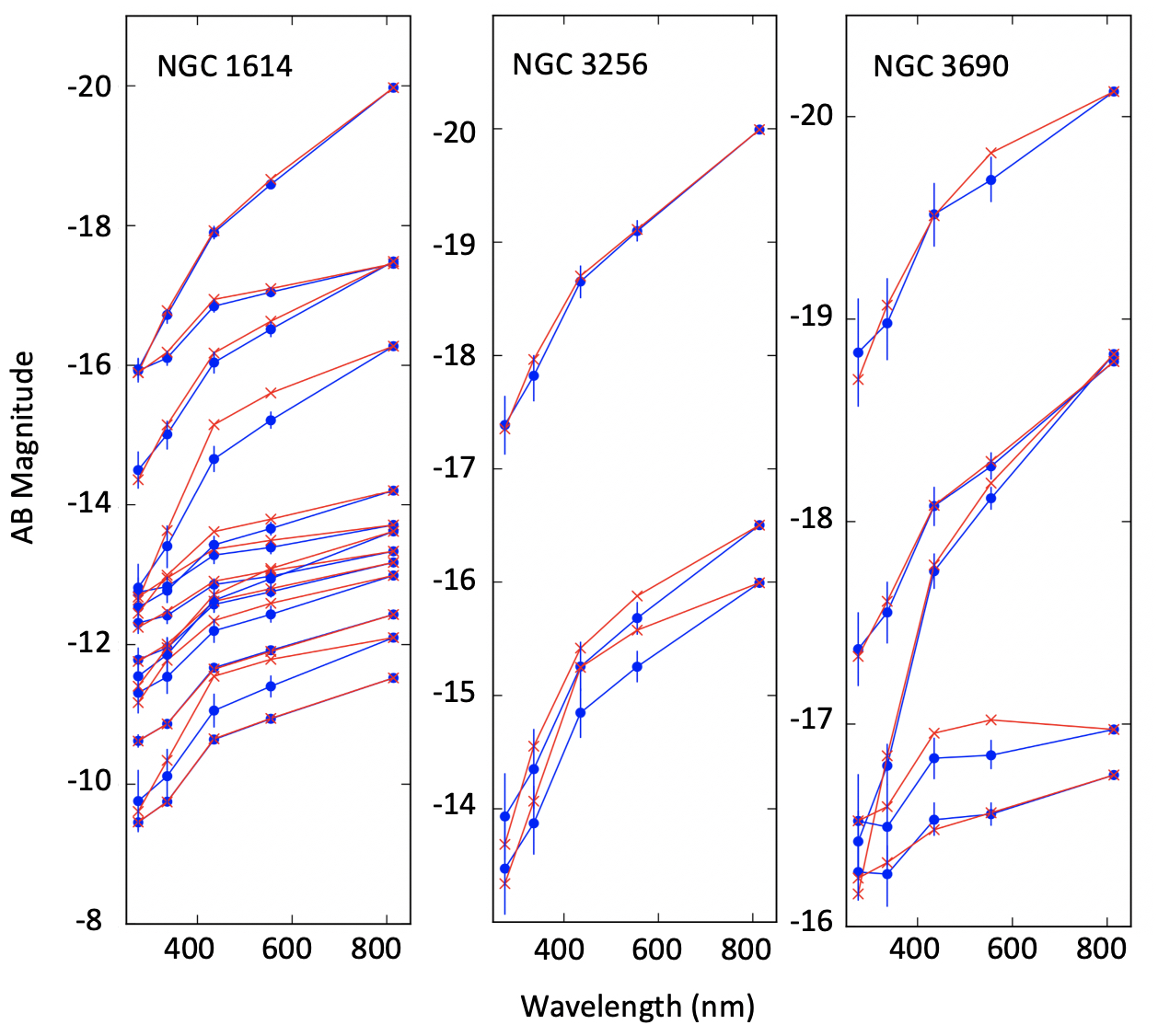}
\caption{Best fits are shown for the clumps for 3 of the galaxies in this study, with blue dots showing the models from the SEDs and red crosses for the measurements. Blue vertical lines indicate uncertainties from the model fits. Some of the fainter ones (at the bottom of the figure) are more poorly fit.}
\label{f14fit}
\end{figure}

Figure \ref{f14fit} shows sample fits for three galaxies. Red crosses represent the
photometric measurements and blue dots with uncertainties are the models. Results
are plotted in AB mags since those were used in the SEDs.

\begin{figure}
\epsscale{1.}
\includegraphics[width=6.5in]{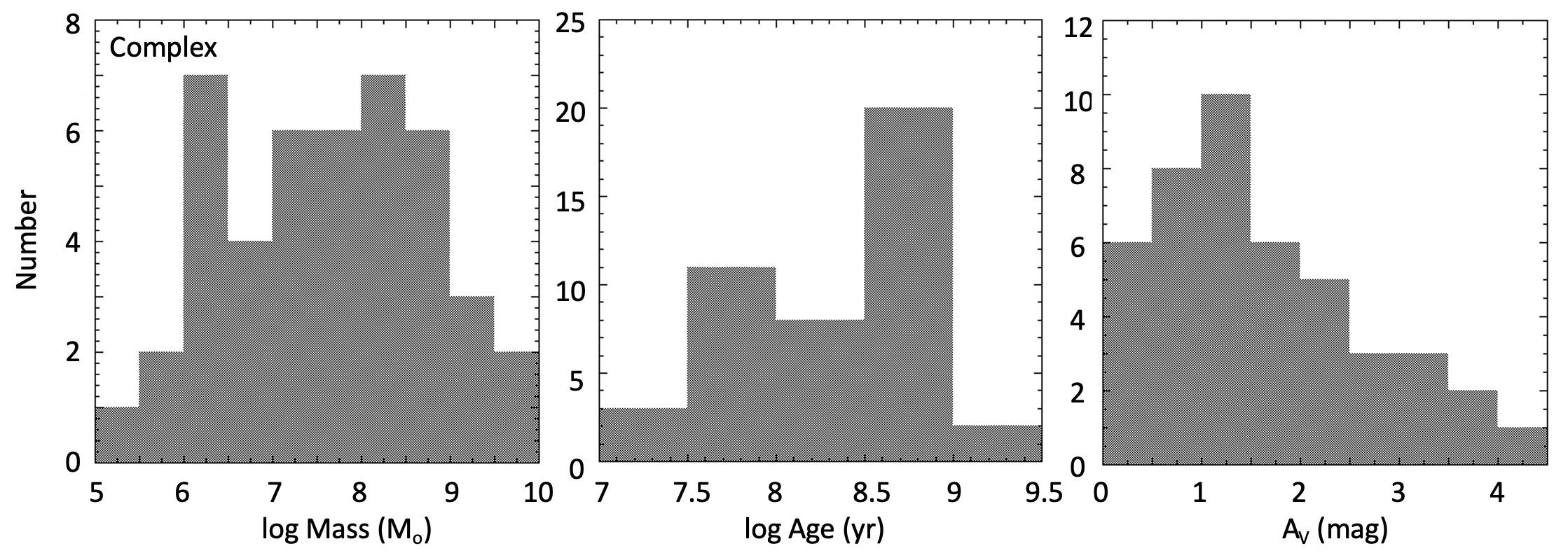}
\caption{Histograms of the masses, ages, and visual magnitudes of extinction of the clumps from all 7 galaxies. The masses are the young star-forming masses, not including the underlying mass of the clump, which would add a factor of $\sim 2$ more mass.}
\label{f15hist}
\end{figure}

The average mass, age, and extinction $A_V$ of the clumps within each galaxy are
tabulated in Table \ref{T2}, and histograms of mass, age, and extinction are shown in Figure \ref{f15hist}. We omitted 7 poorly-fit clumps (3 in Arp 220, 4 in NGC 2623) from the averages and histograms because their mean squared difference between the observed and modeled colors was greater than 1 (in units of mag$^2$).The left-hand panel of Figure
\ref{f15hist} shows a histogram of the young stellar masses in all the clumps
(i.e., not including the underlying component, which would add approximately a
factor of 2 to the clump mass). The log masses in $M_\odot$ range from 5.2 to
9.6, with an average of $7.6\pm 1.2$.
The largest clumps are probably the nuclei of the merging galaxies. For example,
the most massive clump, which is in the center of Arp 220, has a dynamical mass
of $1.5\times 10^9 M_{\odot}$ \citep{wheeler}.

As a check on some individual derived masses, we note that MUSE VLT H$\alpha$
observations of Haro 11 yield a dynamical  mass of $\sim10^8 M_{\odot}$ for the
largest clump \citep{menacho}, compared with our value of $8.15\times 10^7
M_{\odot}$. For the other two clumps, \citet{adamo} derive masses of $8.35\times
10^6 M_{\odot}$ and $1.36\times 10^7 M_{\odot}$ based on point source photometry of
HST observations at the peak brightness, compared with our estimates of $7.76\times
10^7 M_{\odot}$ and $2.29\times 10^7 M_{\odot}$, respectively. Our clump
boundaries are much larger than those in \citet{adamo} since we defined the
extended clumps rather than the peak sources. We derived slightly older
estimated ages since the boundaries extended beyond the brightest star forming site,
but similar extinctions (ours was 1.0 $A_V$ mag compared with their 1.2 mag for
their Knot B).

Our derived log ages in years (shown in the middle panel of Figure
\ref{f15hist}) range from
7.3 to 9, with an average $8.4\pm 0.6$.  The fitted extinctions in $V$-band (shown in the right
panel of the figure) range from 0.2 to 4.2 mag (for a dusty central region in Arp
220), with an average of $1.6\pm 1.0$. Arp 220 and NGC 2623 have the oldest clumps.

The clumps encompass extreme star formation. As an approximate measure of the
average clump star formation rate SFR in $M_{\odot}$ yr$^{-1}$, we take the young
stellar clump mass divided by the age. Then the star formation rate per unit
area, $\Sigma_{SFR}$ in $M_{\odot}$ yr$^{-1}$ kpc$^{-2}$, is given by the SFR
divided by the area of the clump converted to kpc using the values in Table
\ref{T1}. Table \ref{T2} lists the average log(SFR), and the log of the specific
star formation rate, sSFR, taken to be the SFR divided by the total mass in the
clump, including the underlying component plus the new star formation.

\begin{figure}
\epsscale{1.}
\includegraphics[width=4.in]{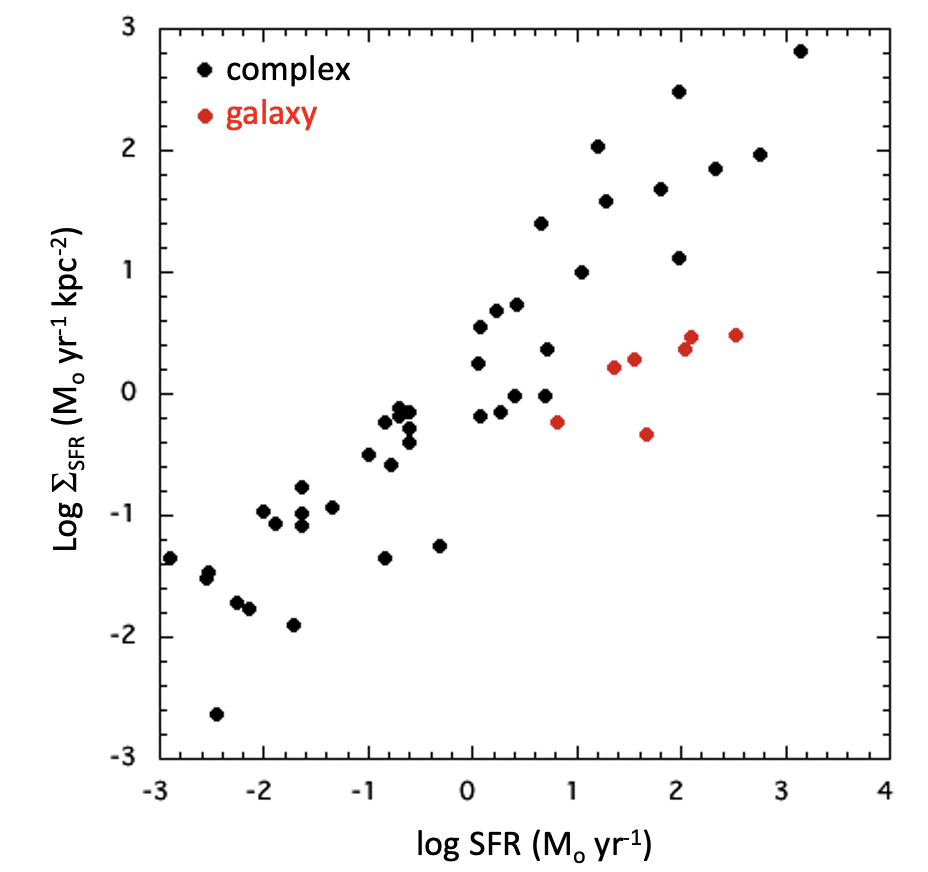}
\caption{Star formation rate per unit area versus star formation rate for the clumps (black dots) and their host galaxies (red dots).}
\label{f16sfr}
\end{figure}

A plot of $\Sigma_{SFR}$ versus SFR is shown in Figure \ref{f16sfr}. The black dots
represent the clumps measured here, and the red dots represent the whole
galaxies \citep{chandar}. The galaxies fall in the realm of other LIRGS. For a
given SFR, the value of $\Sigma_{SFR}$ is about 25$\times$ higher for the clumps
than for each galaxy as a whole. The log sSFR ranges from -9.0 to -7.6, with an
average of $-8.7 \pm 0.4$; average sSFR for clumps within each galaxy are listed
in Table \ref{T2}. These values are similar to those for the ULIRGs in
\citet{u2012} (table 11), which have an sSFR for the whole galaxy ranging from -8.9
to -10.0, with an average of $-9.1\pm 0.4$.    The surface densities of the clumps
in our sample in $M_{\odot}$ pc$^{-2}$ average 1.97$\pm 0.72$ in the log (averages
in each galaxy are in Table \ref{T2}), so are about 100 $M_{\odot}$ pc$^{-2}$.

\begin{figure}
\epsscale{1.}
\includegraphics[width=6.5in]{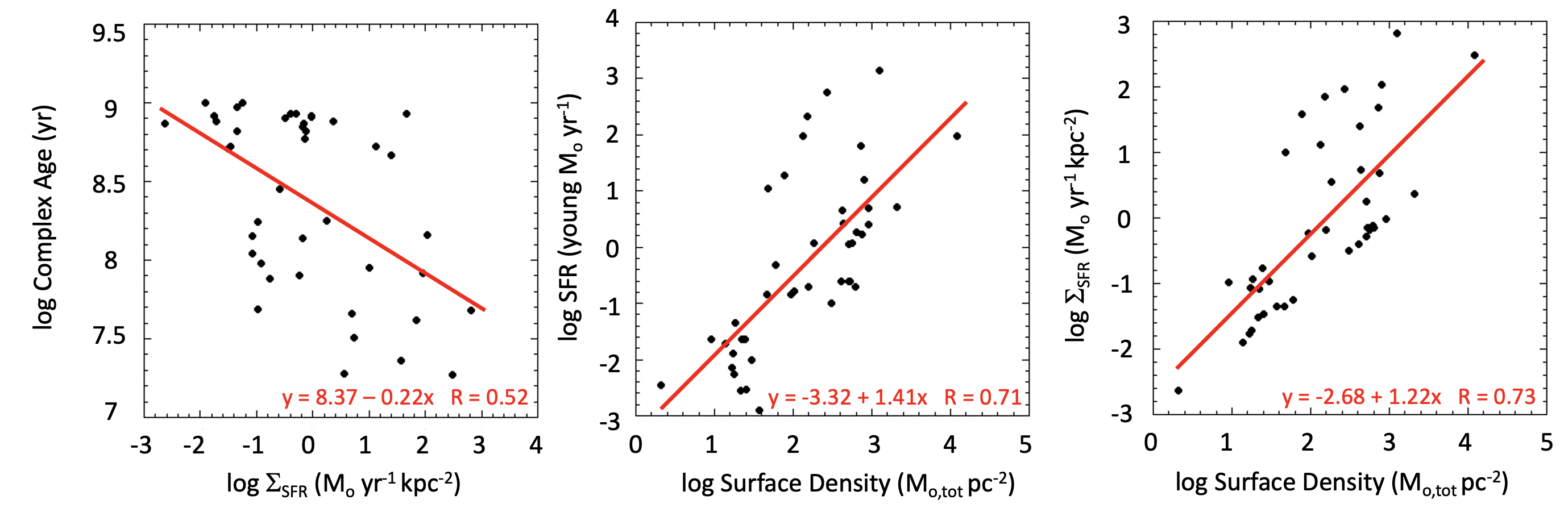}

\caption{(left) Clump age versus star formation rate per unit area; (middle) clump star formation rate versus total stellar surface density; (right) clump star formation rate per unit area versus total stellar surface density.}
\label{f16aSFRdens}
\end{figure}

The clump age as a function of $\Sigma_{SFR}$ is shown in the left panel of
Figure \ref{f16aSFRdens}, along with the curve fit. There a very weak trend of younger clumps having a
higher $\Sigma$. The SFR as a function of total stellar surface density are shown
in the middle panel along with the curve fit. The star formation rate is higher in denser regions.  The
right panel shows a plot of $\Sigma_{SFR}$ versus total stellar surface density along with a curve fit,
also showing a correlation. The range of $\Sigma_{SFR}$ for the clumps is
consistent with the values for the high surface density clumps measured by
\citet{zara} in their large sample of local interacting galaxies (distances up to 140 Mpc). Their figure 4 shows a
linear fit with a broad spread. Our clumps fall on a line displaced upward from
their fit by a factor of $\sim100$, overlapping with the highest star formation
rate per unit area clumps in their sample. This is reasonable, since our
galaxies are mergers and theirs are typically less extreme interactions.

 \subsection{Comparison with high z clumps}
Star-forming clumps in the GEMS, GOODS, and UDF fields were analyzed with the same methods as in this paper, 
so allow a direct comparison of their properties.  
The restframe B magnitudes of clumps
 in interacting galaxies in the GEMS and GOODS fields \citep{elm07a} are
very similar to the local clumps, ranging from -14.5 to -17.5 mag,  while the restframe (B - V) colors of
clumps in clumpy galaxies studied by \cite{elm09a} ranged from -0.8 to 1.8 (the
latter corresponding to central bulge-like clumps), with the majority of the
clump colors between 0 and 1. Thus, the high z clumps are very similar to the
local clumps in brightness and color. 

Clumps in the GEMS and GOODS fields had
log ages decreasing from 9 to 8 from redshifts 0.5 to 1.5, and log masses ranging
from about 7 to 8.6, averaging about 8 across all redshifts from 0.5 to 1.5
\citep{elm09a}. The log surface densities in units of $M_\odot$ pc$^{-2}$ of these
clumps ranged from about 1.5 to 2. They were typically 1 kpc in size, which is
resolved at these redshifts. For ten UDF clumpy galaxies from redshifts 1.6--3, the
clumps averaged a log mass of 8.75 and log age of 8.4; their average log SFR was
0.32 \citep{elm05}, which is essentially the same as the average for the clumps
in the present sample. \citet{elm09b} considered over 2100 clumps in over 400
chain, clump cluster, and spiral galaxies in the UDF out to redshift z=4. The
clumps included star-forming clumps as well as bulge-like clumps; beyond z=5,
the masses averaged $10^8-10^9 M_{\odot}$. Most galaxies showed no signs of
interaction, but a small number showed tidal-like features suggesting mergers.

The local clump diameters are
similar to the kpc-size high redshift clumps in the previously cited
samples. The average diameters for the clumps in each galaxy are listed in Table
\ref{T2}.

\cite{dessuages18} determined the mass function of 194 clumps in galaxies at redshifts from
1 to 3.5 that are seen in deep HST images. They derived a slope of $\sim -1.7$ for
linear intervals of mass, for log mass $> 7.3$. To compare our clumps with theirs, we consider only the higher mass 18 clumps in our sample, with log mass $> 8$. We derive a slope of $-1.4$ for these clumps, which is similar considering our small sample size.

Clumps in clumpy galaxies in the Cosmic Assembly Near-infrared Deep Extragalactic
Legacy Survey \citep[CANDELS;][]{grogin,koek} field were identified by
\citet{guo15}, and their masses were measured by \citet{guo18}. These studies
included nearly 3200 clumps from 1270 galaxies, divided by redshift bins from z=0.5
to 3 and by host galaxy log mass from 9.0 to 11.4. Their figure 4 shows that the
clump masses scale with the galaxy masses, and clumps that had a higher fraction of
UV luminosity relative to the galaxy were slightly more massive than less UV-bright
clumps. Within a given galaxy mass, the results were essentially independent of
redshift bin. For galaxy log masses from 9.0--9.8, the UV-bright clump log masses
ranged from about 6.5 to 9.5, with the majority between 8.0 and 8.5. For galaxy log
masses from 9.8 to 10.6, the majority of the clump log masses were between 8.6 and
9.4. Less UV-bright clumps were factors of 10 to 100 times less massive. These
results are consistent with the local clumps measured in this paper. Clumps in an
additional 6 galaxies at $z \sim 2$ were studied by \citet{fs11} with HST infrared
images and Very Large Telescope (VLT) spectroscopy. They find clump ages ranging
from 50 Myr to 2.75 Gyr.

Evidently, the clumps measured in the present study have a range of masses, ages, sizes,
star formation rates, and surface densities that are similar to high redshift
clumps. This is reasonable, since the turbulent conditions that formed massive clumps at high z apply also to at least the central regions of local interacting and merging galaxies. 
Therefore, these local clumps can be used to probe their substructure.

\section{Clumps and star clusters}
\label{star}

A catalog of compact star clusters for the galaxies in our sample was compiled by
\cite{whit21}. The cluster catalogs were constructed using the DAOPHOT software as
implemented in IRAF. An aperture radius of 2 pixels was used with a sky annulus
between 7 and 9 pixels. A training set of isolated, relatively high S/N clusters
was used to determined aperture corrections for the clusters for the F435W, F555W,
and F814W filters. In most cases, the S/N for the F336W and F275W filters was too
low to be measured directly. In these cases, an offset based on a stellar PSF and
normalized to the F555W filter was used. In cases where both could be measured, the
agreement was roughly 0.2 tenths. An example of typical aperture corrections (i.e.,
for Arp 220) from the 2 pixel aperture to infinity was 1.097, 0,991, 0.821, 0.741,
0.913 for F275W, F336W, F435W, F555W, and F814W, respectively.

At the distance of these galaxies, it is difficult to separate stars from clusters based on resolution in most cases. For this reason, normal manual classification techniques are not being pursued (i.e., unlike for LEGUS \citep{calz} or PHANGS-HST \citep{lee}). Instead, we are using the fact that the brightest stars are generally $\sim M_V = -9$  \citep{hump}, and defining anything brighter than this to be a cluster. Objects fainter than this limit are not included in this paper. We will revisit this and related topics in future papers from the CCDG project.

\begin{figure}
\epsscale{1.}
\includegraphics[width=6.5in]{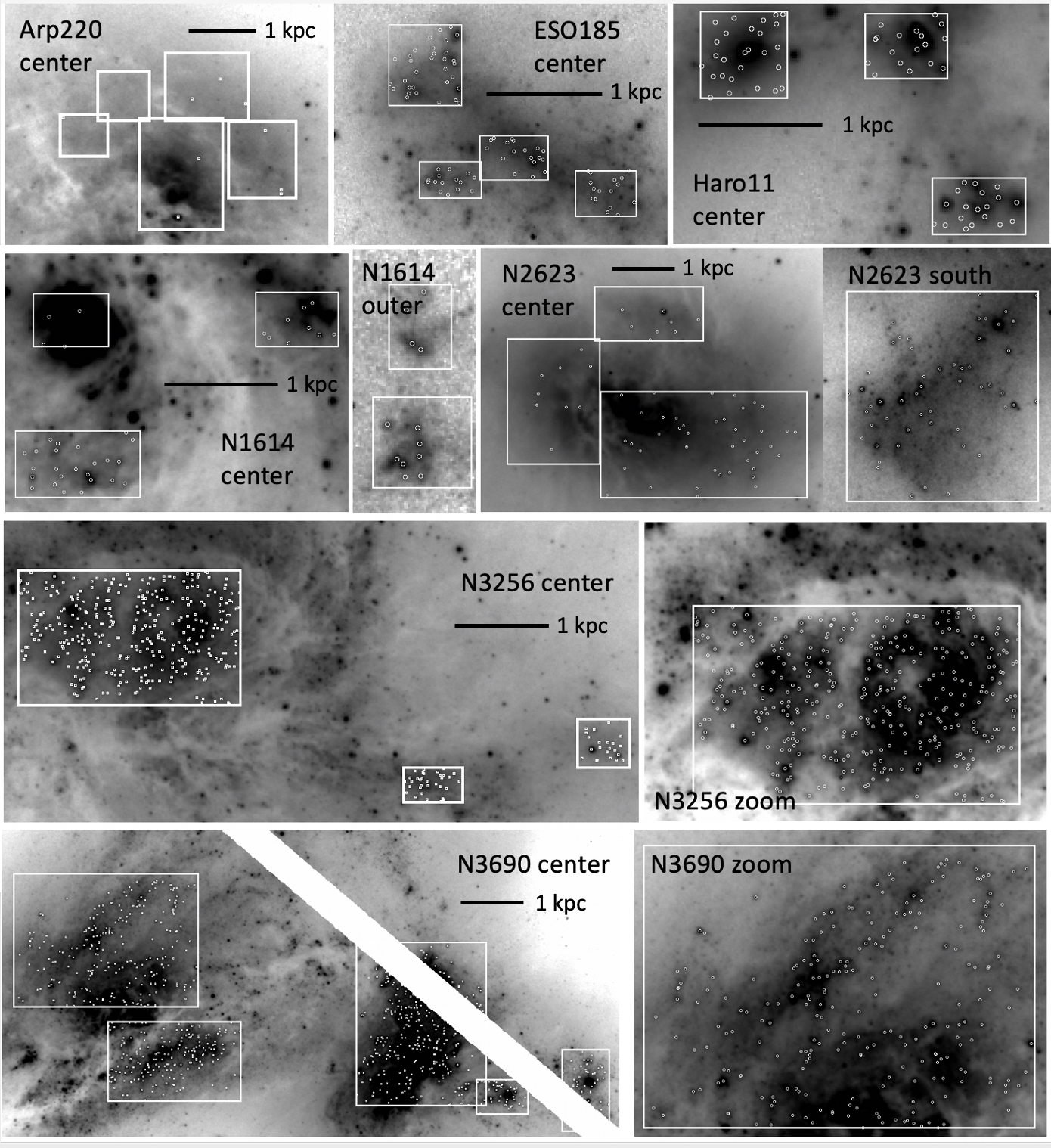}
\caption{Boxes outline some of the clumps in each galaxy, while open circles indicate the locations of clusters within each clump. The images are logarithmic stretches in the F814W filter. Physical scales of 1 kpc are indicated by a solid black line.}
\label{f17over}
\end{figure}

If the local clumps are illustrative of clumps at high redshift, then the
clusters inside these local clumps might be illustrative of clusters inside high
redshift galaxies as well. A total of 1596 clusters brighter than $M_V = -9$ are
within the clumps. The number of clusters inside clumps in each galaxy is
listed in Table \ref{T2}. Figure \ref{f17over} shows the local clusters (circles)
that are within the clump boundaries (rectangles) on the logarithmic stretches of
the F814W images of our galaxies.  Arp 220 is anomalous in having very few clusters
within the clumps, whereas NGC 3256 and NGC 3690 have hundreds. As the figure
illustrates, the distributions of clusters within the clumps are fairly uniform,
although the concentration of clusters increases in regions within a clump where
there are brighter arcs and substructure, such as in NGC 3256 and NGC 3690.

\begin{figure}
\epsscale{1.}
\includegraphics[width=6.in]{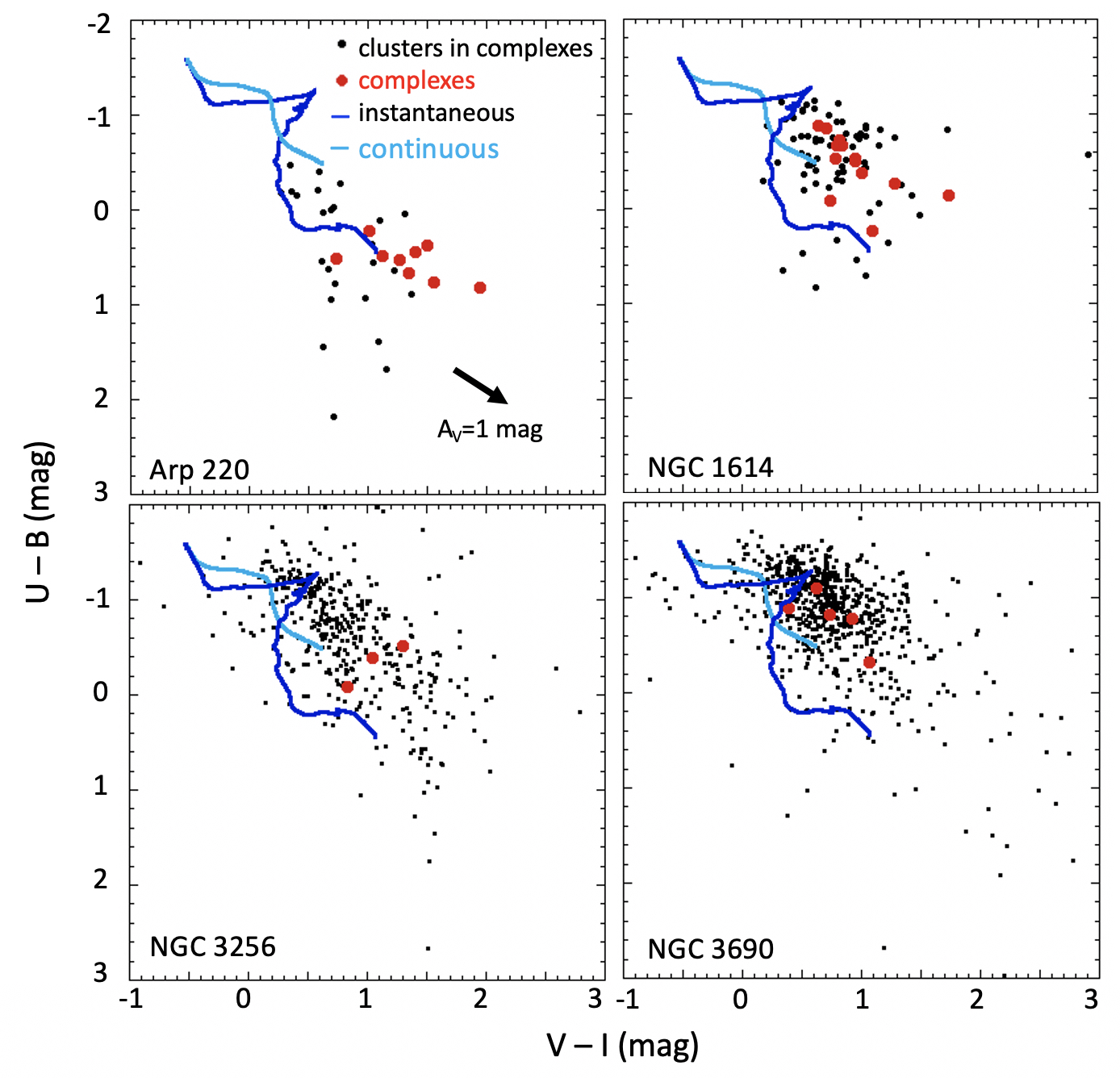}
\caption{Color-color diagram of (U-B) versus (V-I) in Vega mag for clusters and clumps in 4 galaxies. Black dots represent star clusters brighter than $M_V = -9$ that are within a clump, and red dots represent the clumps. These are the observed colors, uncorrected for extinction. Dark blue lines represent the Charlot-Bruzual evolutionary models for solar metallicity, from log age = 5 in the upper left to 10.3 in the lower right, for instantaneous star formation, so these are appropriate for the clusters. The broad spread of clusters indicate a broad range of ages, consistent with ongoing star formation in the clumps rather than a single burst. The light blue lines represent the Charlot-Bruzual models for continuous star formation, as a guide. The clumps were modeled with a combination of the continuous and instantaneous models, as described in the text.}
\label{f12cc}
\end{figure}

Figure \ref{f12cc} shows a color-color plot of (U-B) versus (V-I) in Vega mag for
clusters (black dots) and clumps (red dots) in the four galaxies for which B
observations were available: Arp 220, NGC 1614, NGC 3256, and NGC 3690. These are
the observed colors, not corrected for extinction; a reddening line is drawn to
indicate one magnitude of extinction in V band. The dark blue line is the
evolutionary line from the \citet{bruz} models for a single burst, solar
metalliicity, appropriate for comparison with the star clusters. For reference, the
light blue line is the evolutionary model for continuous star formation. As
described above, the clumps were modeled with a two-component fit using a
combination of the continuous plus instantaneous models. The clumps are
typically redder than the clusters within them, reflecting the underlying older
stars in the clumps. The broad range of colors for the star clusters is
consistent with their formation over a wide range of time, rather than in a single
burst within the clumps.

\begin{figure}
\epsscale{1.}
\includegraphics[width=6in]{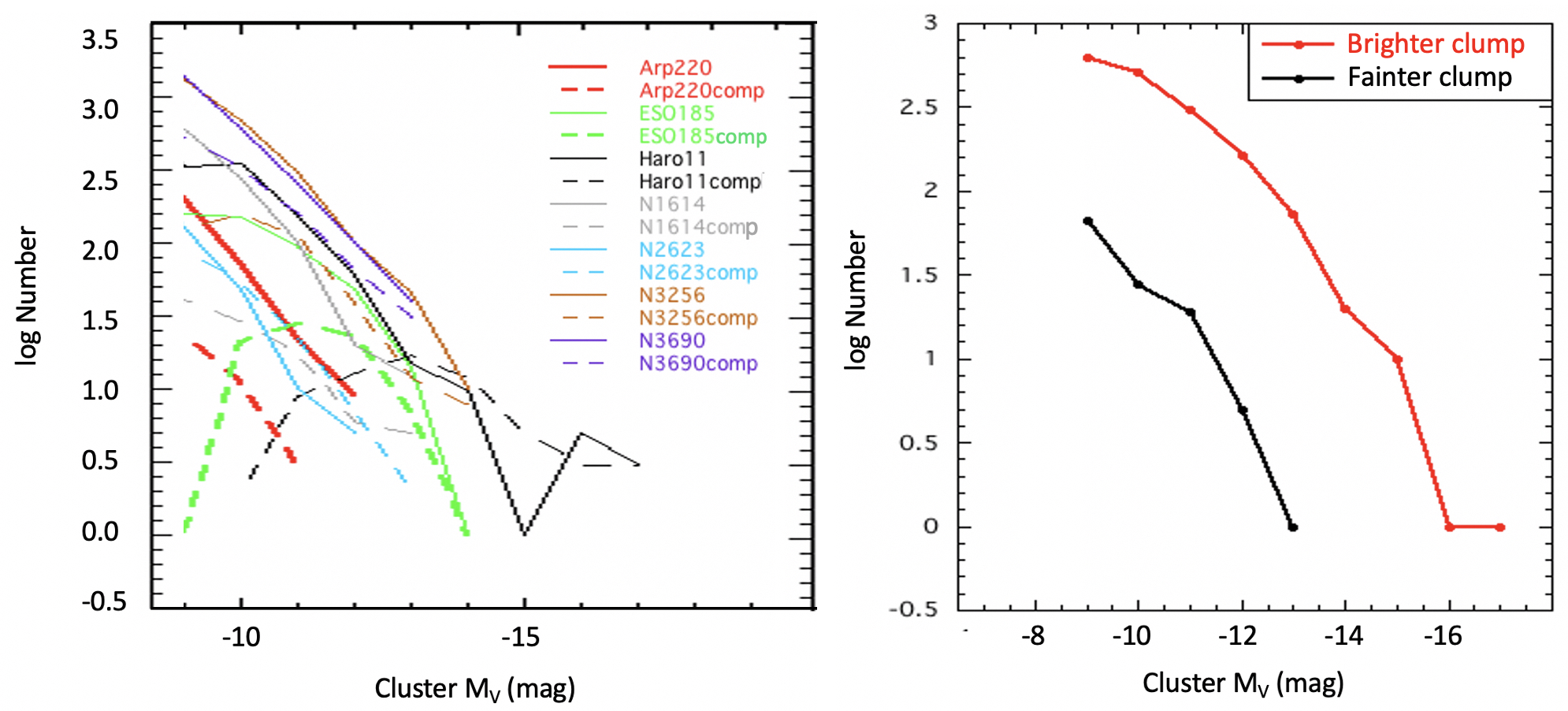}

\caption{(left) Histogram showing the log of the number of clusters as a function of cluster absolute magnitude M$_V$ color-coded by the 7 galaxies. Solid lines are for all the clusters in a galaxy, while dotted
lines are only for clusters that are in clumps. The distributions all have about the same slope.
(right) Clumps in galaxies ESO 185, Haro 11, NGC 1614, NGC 3256, and NGC 3690, which all span
similar ages, are divided into brighter (red line) and fainter (black line) surface brightnesses, with the mid-point being 19 mag arcsec$^{-2}$ in I band. The log of the number of clusters within the clumps is shown versus the cluster
absolute V magnitude. The distributions have the same power law slope. }
\label{f15clushist}
\end{figure}

The magnitude distribution function for clusters is shown in the left-hand panel
of Figure \ref{f15clushist}, color-coded for each galaxy. Solid lines show the
distribution for all the clusters in a given galaxy, while dotted lines show only
the clusters that are within clumps in that galaxy.

The clusters inside clumps are brighter than the clusters outside clumps,
which is likely a size of sample effect \citep[e.g.,][]{whit07}. Overall, the
slopes of the distributions look similar from one galaxy to the next, $\sim -0.4$
on this plot, which has magnitude intervals on the abscissa; this corresponds to a
slope of $-1$ on a similar plot with the base10 log of luminosity function on the
abscissa. In that case the luminosity distribution function in logarithmic
intervals is $N(L) d \log L = L^{-\alpha} d \log L$ for $\alpha = 1$. This is
equivalent to the typical slope for compact star clusters measured elsewhere
\citep{krumholz19}, and is the expected power law function for clusters in a
hierarchical distribution of stellar groupings \citep{elmegreen97}. The slope is
consistent with that for clusters in the Antennae merger \citep{{whit99},{whit}}
and for clusters in 22 LIRG galaxies, including some in our sample, NGC 1614, NGC
2623, NGC 3256, and NGC 3690 \citep{lin}. The left panel of Figure
\ref{f15clushist} also shows that the slopes of the cluster magnitude distributions are about the
same for clusters inside and outside clumps, so the clump environment does not
affect the slope.

There is a possible but uncertain slight steepening of the slope at the bright end of the cluster
distribution function for some of the galaxies.
\citet{adamo11} found a steepening at the bright end of the cluster
function in the dwarf merger galaxy ESO185--IG13 and suggested steepenings
like this in other cluster functions too. However, \citet{mok19} found that the cluster mass function in several galaxies, including NGC 3256 (which is in our sample),
was well fit by a power law, with no bends or breaks.
\citet{mok} used ALMA archival CO data to produce a catalog of giant molecular
clouds (GMCs) in NGC 3256; the GMCs are spread across its central
regions. Both the GMCs and young clusters within them show power law distributions.

The right-hand panel of Figure \ref{f15clushist} plots the magnitude distribution for clusters in two bins of surface brightnesses (red line for brighter and black line for fainter than 19 mag arcsec$^{-2}$) to examine whether the cluster functions differ between the two cases. Since Arp 220 and NGC 2623 have older clumps than the other
galaxies, they were not included in this figure. The other galaxies, ESO 185-IG13,
Haro 11, NGC 1614, NGC 3256, and NGC 3690 have clumps spanning a
similar range of younger ages so they are combined. The average I-band surface
brightness of the clumps in these galaxies is 19 mag arcsec$^{-2}$. The plot
divides clumps into those brighter than average (shown in red) and fainter than
average (shown in black). For each subset, the number of all clusters within all
clumps is shown as a function of the absolute magnitude M$_V$ of the cluster.
The distributions are very similar for the brighter and fainter clumps, although
of course the brighter clumps contain more clusters.

The integrated magnitude for all the clusters in a clump was compared to the
magnitude of the clump in different filters. There was wide variation for
clumps with a galaxy. Overall, clumps were on average 0.1 mag
brighter in NUV (F275W) than the integrated cluster light for Haro 11, about 1
mag for NGC 1614, NGC 3256, and NGC 3690, 3 mag for ESO 185-IG13 and NGC 2623, and
5 for Arp 220. Thus, the clusters
contribute about 90\%, 40\%, 6\%, and 1\% of the NUV clump light for these
galaxies, respectively.

\section{Tidal tails}
\label{tail} The galaxies in this sample all show multiple tidal tails, shells and tidal debris. In what follows, these features are all referred to as ``tidal tails," including
well-defined tails such as in NGC 2623  as well as more diffuse tidal debris or shells such as in NGC 3256, just as was done in the GEMS and GOODS study \citep{elm07a}. ESO 185-IG13, Haro 11, and NGC 3690
have short tails and NGC 3690 is a mid-stage merger \citep{lin}.  NGC 2623 shows
two long tails with a few small wisps; it is thought to be in a more advanced
merger state \citep{{evans},{lin}}. NGC 2623 was also found to have two intense
star formation periods from less than 140 Myr to 1.4 Gyr ago \citep{cor}. Arp 220
has short looping structures and may be a late stage merger with counter-rotating
nuclei \citep{barc}. NGC 1614 and NGC 3256 have similar multiple looping
structures, NGC 3256 is a late-stage merger \citep{lin}. 

\begin{figure}
\epsscale{1.}
\includegraphics[width=5.in]{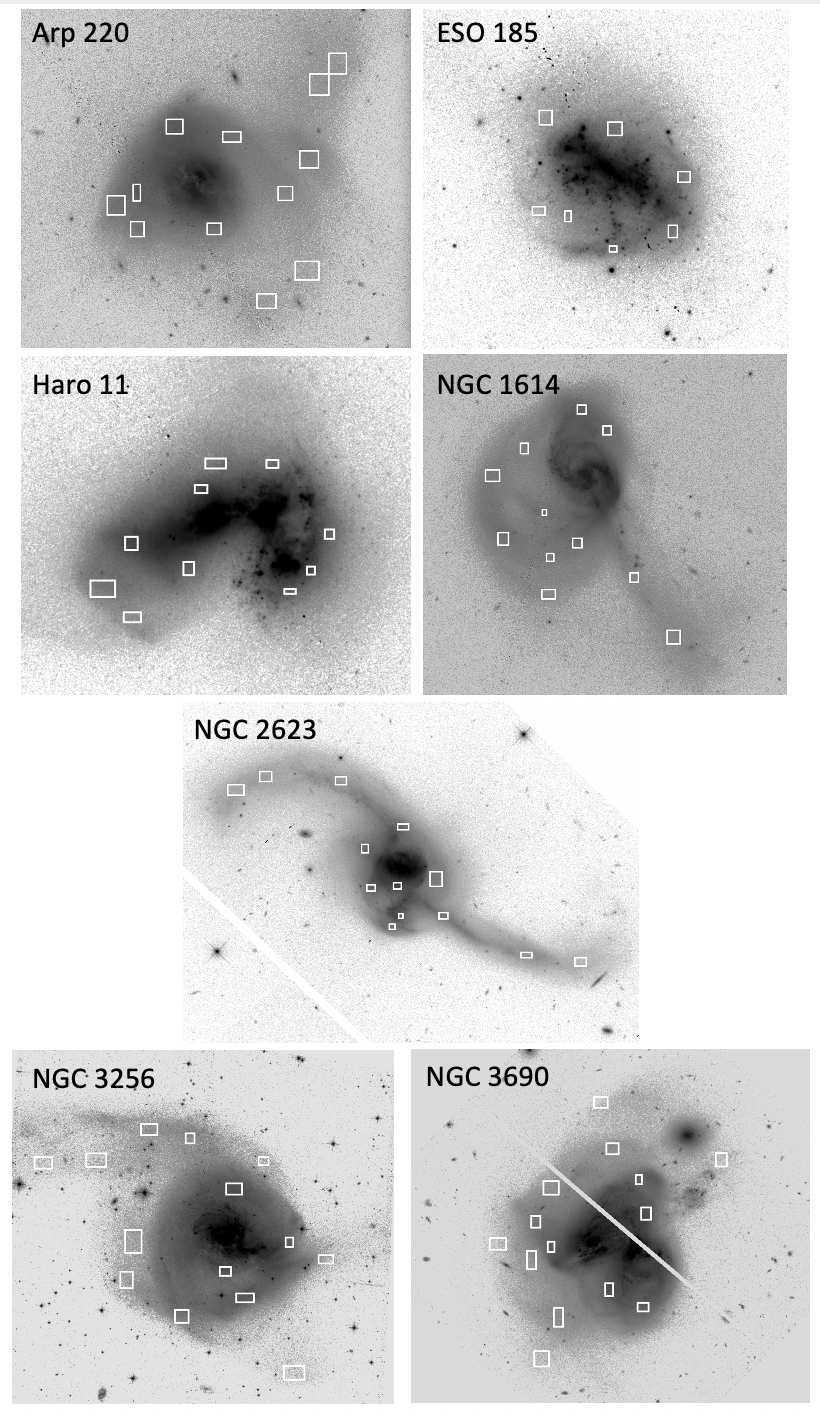}
\caption{Tail locations where surface photometry was done are indicated by boxes on the logarithmic stretches of the F814W images for each galaxy. See Figures 1--4 for physical scales.}
\label{f9}
\end{figure}

In order to compare the tidal tails to those in high redshift galaxies, we sampled the tails in several different positions to get their average properties. These positions avoided obvious star clusters and clumps. 
Figure
\ref{f9} shows boxes for the 81 positions in which surface photometry was done, using the IRAF task {\it imstat} as was done for the clumps.
Table \ref{T3} lists the average surface brightnesses and colors of the tidal tails in each galaxy. Both of these quantities are similar in ESO 185 and Haro
11, which are morphologically similar also. NGC 2623 has the highest surface
brightness arms. The surface brightnesses are similar for Arp 220 and NGC 3690 and
for NGC 1614 and NGC 3256. Average values for the features measured in each galaxy are also listed. The average (V-I) color is about 1, and ranges from
about 0.5 to 1.5 mag independent of surface brightness.

\citet{mullan} studied tidal tails in a diverse sample of 17 local interacting galaxies
with HST observations, using the same method as in this paper to measure several random positions within the tidal arms. 
There, the V-band tail surface brightnesses averaged 24
arcsec$^{-2}$ and ranged from 22 to 25.5 arcsec$^{-2}$. The average (V-I) color was
1, ranging from 0.6 to 1.4, so their properties are similar to those of the tails and debris
in the current sample, although there were not any tails in the Mullan study as bright as the brightest
tails here.

\begin{figure}
\epsscale{1.}
\includegraphics[width=4.in]{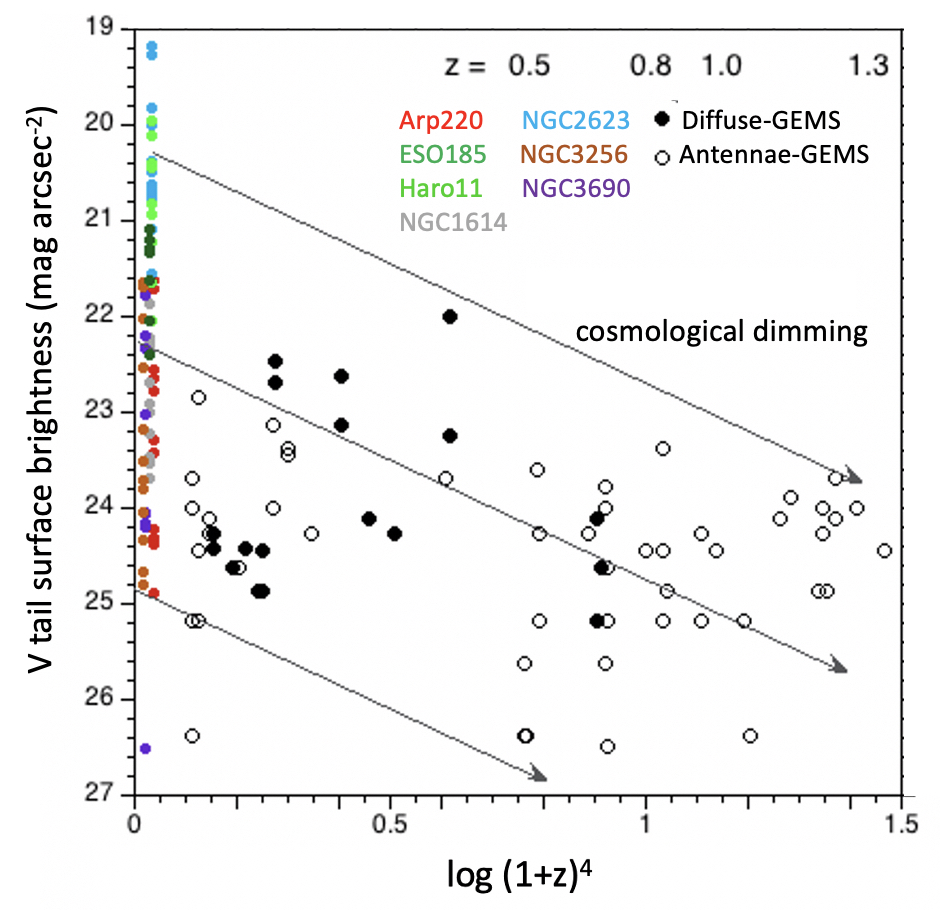}
\caption{V-band tail surface brightness is plotted as a function of log (1+z)$^4$. Corresponding redshifts, z, are indicated along the top x axis. Measurements of tails from the local merger sample are labeled. Tails from GEMS and GOODS galaxies  \citep{elm07a} are marked as diffuse (closed circles) or antennae (open circles); for compactness in the legend, those labeled as GEMS also include GOODS galaxies. The latter used F606W, instead of F555W used in the present sample. The lines indicate cosmological dimming.}
\label{f10z}
\end{figure}

The V-band surface brightnesses of the GEMS and GOODS tails \citep{elm07a} compared
with the galaxies in this sample are shown in Figure \ref{f10z} as a function of
log (1+z)$^4$, since cosmological dimming decreases surface brightness by
(1+z)$^4$. Thus, there is a drop of 2.5 mag for each 1 unit drop in log(1+z)$^4$.
For z=0.5, 0.8, 1, and 2, the corresponding decreases are 1.8, 2.5, 3, and 4.8 mag
arcsec$^{-2}$.The $2\sigma$ surface brightness detection is about 24.5 mag
arcsec$^{-2}$ in the GEMS and GOODS fields. For a z=0.8 galaxy, this would
correspond to a local tail surface brightness of 22 mag arcsec$^{-2}$, which is the
average value for the local galaxies in this sample. The tails in NGC 2623 and Haro
11 are so bright that they would still be visible at z=2, while all but the
brightest patches in Arp 220, NGC 1614, or NGC 3690 would not be observed. Thus,
some galaxies beyond z=1 could be mergers that are not obvious from their tidal
features.

Tidal tails may be observable for only the first few hundred million years in the
early stages of a merger because the surface brightness dims as the stars age and
disperse \citep{mihos}. \cite{hibbard} measured the surface brightness of the
nearby merger, Arp 299, which has an interaction age of 750 Myr. The faintest tail
region they measured goes down to 28.5 mag arcsec$^{-2}$ in the B band. This would
be undetectable at high z using the limits assumed here.

Our redshifted HST images of local mergers show some large star-forming clumps in
the tidal tails. In some galaxies, such clumps may be tidal dwarfs, as studied by
\cite{duc}, \cite{mir} and others, but that does not appear to be the case in the
current sample. Similar tidal clumps were discussed for high-z galaxies
\citep{{elm07a},{zanella19}}.

\section{Dust and gas}
\label{dustgas}
\subsection{Extinction minima}
\label{dust}

Some of the clumps in the local galaxies are outlined by dust
extinction, as evident from the wispy structure occulting the underlying disk in
the HST images.  The old red clumps in the artificially redshifted image of Arp 220 are of
this type, being the inner clear regions of each disk. CO emission in the dusty
region between the two bright lobes is seen at low resolution \citep{sco,barc}
and CO emission centered on the optical lobes is seen at high resolution
\citep{brown}. There are relatively few star clusters in these Arp 220 regions
however, presumably because they are old, and that also accounts for the ISM
clearing around them. Nevertheless a morphology of clumps as artifacts from
extinction minima is possible at high redshift too.

Other galaxies in our sample have thin dusty molecular features like Arp 220. NGC
1614 has a thin dust lane cutting across one of the inner clumps next to a
broader dust lane, as seen in Figure \ref{f7}. Its CO emission observed with ALMA
is centered on the dust feature and encompasses the western bright regions
\citep{konig}. NGC 3256 also has a dust feature cutting the central region (see
Figure \ref{f8} and Figure 2 of \citet{zepf}), although it does not cut through a
clump.

Haro 11, a low metallicity BCD, has no detectable HI or CO, with upper limits of
$\sim 10^8 M_{\odot}$ for each component \citep{berg}. Its ionized and diffuse gas component is
  $5.8\times10^8 M_{\odot}$  based on mid- and far-infrared
fine-structure cooling lines observed with the Spitzer Infrared Spectrograph (IRS)
and Herschel Photodetector Array Camera and Spectrometer (PACS) by  \citet{corm}, who describe the
photodissociation region and compact HII regions associated with the three main
clumps.

NGC 2623 shows CO emission extended across the central region \citep{brown}. It has
a dust feature cutting across the middle clump, as evident particularly in the
F275W image of Figure \ref{f6} (see also \citealt{adamo}) but the clump still
appears as a single clump when artificially redshifted.

Some bright regions at high redshift could be extinction minima too, appearing as clumps because dust obscures other parts of the extended emission. \citet{tacc08}
observed a large CO cloud between the visible star-forming clumps in the
submillimeter galaxy N2850.4 at z=2.39, suggesting the bright clumps are not
associated with the densest gas in that case.

These observations suggest that some apparent clumps viewed at high redshift could
be only extinction minima, but this is actually rare in our redshifted local
sample. Only Arp 220 shows clear evidence for it, and what appears through the
extinction is not a star-forming region but two old inner disks of the colliding
galaxies.

\subsection{Gas similarities}
\label{gas}

Local mergers can also be gas-rich with high surface densities, like high redshift
disk galaxies. In our sample, a molecular gas mass of $10^{10} M_{\odot}$ has been
detected in the central region of Arp 220 \citep{sco} at high density
\citep{brown}, with gas surface densities of 2.2--4.5$\times10^5 M_{\odot}$
pc$^{-2}$ \citep{barc}. NGC 3256 has a molecular nuclear disk with a gas density $>
10^3 M_{\odot}$ pc$^{-2}$ \citep{saka}, with HCN and HCO$^+$ outflows \citep{mich}.
Dense gas tracer isotopes of CO and CN were detected in NGC 1614 \citep{konig},
isotopes of CO in NGC 2623 \citep{brown}, and HCO$^+$, HCN, and other dense gas
tracers in NGC 3690 \citep{jiang}. CO has been detected with comparable surface
densities ($> 500-700 M_{\odot}$ pc$^{-2}$)  in $z\sim 2$ galaxies by
\citet{tacc08,tacc} and others. For the local mergers, the high gas surface
density is the result of in-plane accretion, tidal compression, and fast
compressive turbulence \citep{renaud14}. For high redshift disks, the high gas
surface density is presumably from a combination of a large accretion flux
continuously processed into stars \citep{bouche10}, plus occasional mergers
\citep{consel}.

Local mergers and high z disks also have high gas velocity dispersions. For
example, high CO velocity dispersions with no ordered rotation suggest mergers in
four $\sim$2 submillimeter galaxies \citep{tacc08}, and high H$\alpha$ dispersions
suggest mergers in two UV-bright galaxies \citep{fs06}. Local interacting galaxies
can have high gas velocity dispersions also, such as 50 km s$^{-1}$ FWHM or more in
HI \citep{kauf95,kauf12}, and local mergers have high gas dispersions too, such as Arp 220 at 300 km s$^{-1}$ \citep{sco}.

\section{Conclusions}
\label{conc}

HST images of 7 strongly interacting and merging disk galaxies in the local universe
observed in the CCDG sample have been blurred, dimmed, and re-pixelated to match the observing conditions at
high redshift.  With these changes, the local galaxies appear similar to high
redshift star-forming galaxies: both are clumpy, and the clumps have about the
same range of physical size, mass, intrinsic surface brightness, age, and star
formation rate at low and high redshift. This is in contrast to clumps in local non-interacting galaxies, which are smaller and less massive than high redshift clumps.
The observed clumps also have the same range of surface density, star formation per unit area,
and specific star formation rate as high redshift clumps.

These similarities, combined with the loss due to cosmological dimming at high redshift of low surface
brightness features seen in the local galaxies, such as tidal tails, suggest that some
clumpy high-z galaxies that look isolated could really be mergers. Other
ambiguities about the characteristics of mergers were discussed in the
introduction.

We also studied catalogued star clusters in the local galaxies and found that the
 clumps contain star clusters with normal luminosity functions. We
infer from this that high redshift clumps contain (unresolved) normal bound clusters also, as a consequence of a hierarchy of star formation.

We thank the referee for helpful suggestions to improve the paper. Based on observations with the NASA/ESA Hubble Space Telescope, obtained at the Space Telescope Science Institute, which is operated by the Association of Universities for Research in Astronomy, Incorporated, under NASA contract NAS5-26555. Support for program number HST-GO-15649 was provided through a grant from the STScI under NASA contract NAS5-26555. This research has made use of the NASA/IPAC Extragalactic Database (NED) which is operated by the Jet Propulsion Laboratory, California Institute of Technology, under contract with the National Aeronautics and Space Administration.

\newpage

\newpage

\newpage

\newpage

\newpage

\newpage

\newpage

\newpage

\newpage
\begin{deluxetable}{lcccccccccc}
\tabletypesize{\scriptsize}
\tablecolumns{11}
\tablewidth{0pt}
\tablecaption{Galaxies}
\label{T1}
\tablehead{
\colhead{Name}&
\colhead{F275W}&
\colhead{F336W}&
\colhead{F435W}&
\colhead{F438W}&
\colhead{F555W} &
\colhead{F814W} &
\colhead{Distance}&
\colhead{Scale} &
\colhead{No. clumps}&
\colhead{No. clusters}
\\
\colhead{}&
\colhead{camera}&
\colhead{camera}&
\colhead{camera}&
\colhead{camera}&
\colhead{camera} &
\colhead{camera} &
\colhead{Mpc}&
\colhead{pc/px} &
\colhead{}&
\colhead{(in clumps)}
}
\startdata
Arp 220	&	WFC3	&	WFC3	&	ACS	&	 \nodata	&	WFC3	&	ACS	&	81.5	&	15.6	&	9	&28	\\
ESO185--IG13	&	WFC3	&	 \nodata	&	 \nodata	&	WFC3	&	WFC3	&	WFC3	&	81.6	&	15.6	&	4	&81		\\
Haro11	&	WFC3	&	 \nodata	&	ACS	&	 \nodata	&	WFC3	&	WFC3	&	90.8	&	17.4	&	3	&65		\\
NGC1614	&	WFC3	&	WFC3	&	ACS	&	 \nodata	&	WFC3	&	ACS	&	69.2	&	13.2	&	13	&72		\\
NGC2623	&	WFC3	&	 \nodata	&	ACS	&	 \nodata	&	ACS	&	ACS	&	80.9	&	15.5	&	13	&116		\\
NGC3256	&	WFC3	&	WFC3	&	ACS	&	 \nodata	&	ACS	&	ACS	&	38.1	&	7.3	&	4	&406		\\
NGC3690	&	WFC3	&	WFC3	&	ACS	&	WFC3	&	WFC3	&	ACS	&	47.3	&	9.1	&	5&828	
\enddata
\tablecomments{D$_{GSR}$ from the NASA/IPAC Extragalactic Database
(NED; https://ned.ipac.caltech.edu/). }
\end{deluxetable}

\begin{deluxetable}{lccccccc}
\tabletypesize{\scriptsize}
\tablecolumns{8}
\tablewidth{0pt}
\tablecaption{Average Properties of Clumps}
\label{T2}
\tablehead{
\colhead{Name}&
\colhead{Diameter}&
\colhead{$\log(Mass)$}&
\colhead{$\log(Age)$}&
\colhead{$A_V$}&
\colhead{$\log(SFR)$}&
\colhead{$\Sigma$}&
\colhead{sSFR}
\\
\colhead{ } &
\colhead{(kpc)}&
\colhead{(M$_{\odot}$)}&
\colhead{(yr)}&
\colhead{(mag)}&
\colhead{($M_{\odot} yr^{-1}$)}&
\colhead{($M_{\odot} kpc^{-2}$)}&
\colhead{($M_{\odot} yr^{-1} kpc^{-2} $)}
}
\startdata	 	
Arp 220	&	1.30	$\pm$	0.45	&	8.74	$\pm$	0.57	&	8.90	$\pm$	0.05	&	2.95	$\pm$	0.90	&	-0.16	$\pm$	0.61	&	2.59	$\pm$	0.48	&	-8.93	$\pm$	0.16\\
ESO185	&	0.60	$\pm$	0.13	&	7.51	$\pm$	0.37	&	8.35	$\pm$	0.43	&	0.52	$\pm$	0.049	&	-0.84	$\pm$	0.12	&	2.17	$\pm$	0.23	&	-8.54	$\pm$	0.34\\
Haro11	&	0.62	$\pm$	0.06	&	7.73	$\pm$	0.32	&	7.48	$\pm$	0.19	&	0.40	$\pm$	0.25	&	0.24	$\pm$	0.17	&	2.60	$\pm$	0.31	&	-7.90	$\pm$	0.24\\
N1614	&	0.45	$\pm$	0.17	&	6.92	$\pm$	1.11	&	8.26	$\pm$	0.49	&	1.63	$\pm$	0.83	&	-1.34	$\pm$	1.3	&	1.87	$\pm$	0.92	&	-8.45	$\pm$	0.42\\
N2623	&	1.30	$\pm$	0.79	&	7.44	$\pm$	1.59	&	8.74	$\pm$	0.42	&	1.76	$\pm$	1.05	&	-1.12	$\pm$	1.62	&	1.44	$\pm$	1.18	&	-8.65	$\pm$	0.62\\
N3256	&	0.75	$\pm$	0.61	&	8.17	$\pm$	0.71	&	8.17	$\pm$	0.50	&	2.06	$\pm$	0.31	&	1.66	$\pm$	1.31	&	2.87	$\pm$	0.24	&	-6.79	$\pm$	0.48\\
N3690	&	1.73	$\pm$	0.85	&	8.13	$\pm$	0.78	&	7.91	$\pm$	0.51	&	1.09	$\pm$	0.47	&	1.88	$\pm$	0.71	&	2.06	$\pm$	0.29	&	-6.56	$\pm$	0.29	
\enddata
\tablecomments{The headers average properties of clumps within each galaxy (with the number of clumps listed in Table \ref{T1}. Mass, age, and extinction are from the SED fits;  mass and age refer to the new star formation in the clump. SFR is the young stellar mass divided by the age. $\Sigma$ is the surface density, from total mass of the clump including the young and underlying component, divided by the area. sSFR is the specific star formation rate, taken as the SFR divided by the total mass. }
\end{deluxetable}

\newpage
\begin{deluxetable}{lccccccccc}
\tabletypesize{\scriptsize}
\tablecolumns{10}
\tablewidth{0pt}
\tablecaption{Tail Properties}
\label{T3}
\tablehead{
\colhead{Galaxy}&
\colhead{NUV}&
\colhead{U}&
\colhead{B}&
\colhead{V} &
\colhead{I} &
\colhead{NUV-U}&
\colhead{U-B} &
\colhead{B-V}&
\colhead{V-I}\\
\colhead{}&
\colhead{mag arcsec$^{-2}$}&
\colhead{mag arcsec$^{-2}$}&
\colhead{mag arcsec$^{-2}$}&
\colhead{mag arcsec$^{-2}$}&
\colhead{mag arcsec$^{-2}$}&
\colhead{mag}&
\colhead{mag}&
\colhead{mag }&
\colhead{mag}
}
\startdata
Arp 220	&	25.35	$\pm$	1.64	&	24.25	$\pm$	0.98	&	23.56	$\pm$	0.91	&	23.2	$\pm$	1.07	&	21.92	$\pm$	0.96	&	1.27	$\pm$	1.82	&	0.69	$\pm$	0.24	&	0.36	$\pm$	0.18	&	1.27	$\pm$	0.14	\\
ESO185--IG13	&	21.84	$\pm$	0.65	&	\nodata			&	22.00	$\pm$	0.54	&	21.57	$\pm$	0.48	&	20.92	$\pm$	0.51	&	\nodata	 	 	&	\nodata			&	0.42	$\pm$	0.07	&	0.65	$\pm$	0.54	\\
Haro11	&	21.23	$\pm$	1.34	&	\nodata			&	21.27	$\pm$	0.69	&	20.84	$\pm$	0.66	&	20.12	$\pm$	0.54	&	\nodata	 	 	&	\nodata			&	0.43	$\pm$	0.06	&	0.73	$\pm$	0.21	\\
NGC 1614	&	25.19	$\pm$	0.59	&	23.62	$\pm$	0.56	&	23.48	$\pm$	0.62	&	22.84	$\pm$	0.61	&	21.66	$\pm$	0.62	&	1.98	$\pm$	0.54	&	0.15	$\pm$	0.16 	&	0.63	$\pm$	0.09	&	1.18	$\pm$	0.07	\\
NGC 2623	&	22.11	$\pm$	1.10	&	\nodata			&	20.84	$\pm$	0.63	&	20.33	$\pm$	0.67	&	19.40	$\pm$	0.72	&	\nodata	 	 	&	\nodata			&	0.52	$\pm$	0.08	&	0.93	$\pm$	0.09	\\
NGC 3256	&	24.94	$\pm$	1.49	&	24.84	$\pm$	1.72	&	24.28	$\pm$	1.28	&	23.22	$\pm$	1.12	&	22.71	$\pm$	1.64	&	1.04	$\pm$	1.49	&	0.80	$\pm$	1.08	&	1.18	$\pm$	0.70	&	0.51	$\pm$	0.60	\\
NGC 3690	&	24.51	$\pm$	0.99	&	24.04	$\pm$	1.52	&	24.02	$\pm$	1.46	&	23.29	$\pm$	1.47	&	22.62	$\pm$	1.65	&	0.79	$\pm$	0.64	&	0.00	$\pm$	0.31	&	0.44	$\pm$	0.52	&	1.29	$\pm$	0.25	\\
Average	&	23.50	$\pm$	2.02	&	24.19	$\pm$	1.31	&	22.94	$\pm$	1.67	&	22.24	$\pm$	1.51	&	21.43	$\pm$	1.63	&	1.19	$\pm$	1.32	&	0.42	$\pm$	0.65	&	0.61	$\pm$	0.47	&	0.94	$\pm$	0.41	
\enddata
\tablecomments{The headers correspond to average surface brightnesses or colors (for the number of tail measurements listed in Table 1) for the different filters: NUV is F275W, U is F336W, B is either F435W or F438W, V is F555W, and I is F814W. }
\end{deluxetable}


\begin{thebibliography}

\bibitem[Adamo et al.(2020)]{adamo20} Adamo, A.,  Hollyhead, K., Messa,M., Ryon, J .E., Bajaj, V. , Runnholm, A., Aalto, S., Calzetti, D.,  Gallagher, J. S., Hayes , M. J.,  Kruijssen , J. M. D., K\"{o}nig, S.,  Larsen, S. S.,  Melinder, J.,  Sabbi, E., Smith, L. J. \& . \"{O}stlin, G.  2020, MNRAS, 499, 3267

\bibitem[Adamo et al.(2010)]{adamo} Adamo, A., \"{O}stlin, G., Zackrisson, E., et al. 2010, MNRAS, 407, 870

\bibitem[Adamo et al.(2011)]{adamo11} Adamo, A., \"{O}stlin, G., Zackrisson, E., \& Hayes, M. 2011, MNRAS, 414, 1793

\bibitem[Barcos-Mu\~{n}oz et al.(2015)]{barc} Barcos-Mu\~{n}oz, L., Leroy, A. K., Evans, A. S., et al. 2015, ApJ, 799, 10

\bibitem[Beckwith, Stiavelli, Koekemoer et al.(2006)]{beck} Beckwith, S., Stiavelli, M., Koekemoer, A., et al. 2006, AJ, 132, 1792

\bibitem[Bergvall et al.(2000)]{berg} Bergvall, N., Masegosa, J., \"{O}stlin, G., \& Cernicharo, J. 2000, A\&A, 359, 41

\bibitem[Blumenthal et al.(2020)]{blumenthal20} Blumenthal, K.A., Moreno, J., Barnes, J.E.,
Hernquist, L., Torrey, P., Claytor, Z. Rodriguez-Gomez, V., Marinacci, F., Vogelsberger, M. 2020, MNRAS, 492, 2075


\bibitem[Bouch\'{e} et al.(2010)]{bouche10} Bouch\'{e} N., Dekel A., Genzel R., Genel S., Cresci G., F\"orster Schreiber N.M., Shapiro K.L., Davies R.I., Tacconi, L. 2010, ApJ, 718, 1001

\bibitem[Bournaud \& Elmegreen(2009)]{bour} Bournaud, F. \& Elmegreen, B.G. 2009, ApJ, 694, L158

\bibitem[Brown \& Wilson(2019)]{brown} Brown, T. \& Wilson, C.D. 2019, ApJ, 819, 17

\bibitem[Bruzual \& Charlot(2003)]{bruz} Bruzual, G., \& Charlot, S. 2003, MNRAS, 344, 1000

\bibitem[Calzetti et al.(2015)]{calz} Calzetti, D., Lee, J.C., Sabbi, E., Adamo, A., Smith, L.J. et al. 2015, AJ, 149, 51

\bibitem[Cava et al.(2017)]{cava} Cava, A., Schaerer, D., Richard, J., P\'{e}rez-Gonz\'{a}lez, P.G., Dessauges-Zavadsky, M., Mayer, L., \& Tamburello, V. 2017, Nat. Astr., 2, 76

\bibitem[Ceverino, Dekel \& Bournaud(2010)]{cev} Ceverino, D., Dekel, A., \& Bournaud, F. 2010, MNRAS, 404, 2151


\bibitem[Chandar(2018)]{chandar} Chandar, R., HST Proposal. Cycle 26, ID. \#15649

\bibitem[Chandar et al.(2021)]{chandar21} Chandar, R., Whitmore, B., Calzetti, D., Lee, J., Cook, D., Elmegreen D., Mok, A., Ubeda, L., \& White, R.,   in preparation

\bibitem[Cibinel et al.(2019)]{cibinel19} Cibinel, A., Daddi, E., Sargent, M. T., et al. 2019, MNRAS, 485, .5631


\bibitem[\'Ciprijanovi\'c et al.(2020)]{cipri20} \'Ciprijanovi\'c, A., Snyder, G.
    F., Nord, B., \& Peek, J. E. G. 2020, Astronomy and Computing, 32, 100390



\bibitem[Conselice et al.(2003)]{conselice03} Conselice, C.J., Bershady, M.A., Dickinson, M., Papovich, C. 2003, AJ, 126, 1183

\bibitem[Conselice(2014)]{consel} Conselice, C. J. 2014, ARA\&A, 52, 291

\bibitem[Cormier et al.(2012)]{corm} Cormier, D., LeBouteiller, V., Madden, S.C., Abel, N., Hony, S. et al. 2012, A\&A 548, A20

\bibitem[Cortijo-Ferrero et al.(2017)]{cor} Cortijo-Ferrero, C., Gonz\'{a}lez Delgado, R. M., P\'{e}rez, E., Cid Fernandes, \& Garc\'{i}a-Benito, R. 2017, A\&A, 607, A70

\bibitem[Cowie, Hu, \& Songaila(1995)]{cowie} Cowie, L., Hu, E., \& Songaila, A. 1995, AJ, 110, 1576

\bibitem[Darg et al.(2010)]{darg10} Darg, D. W., Kaviraj, S., Lintott, C. J., et al, 2010, MNRAS, 401, 1043


\bibitem[Dav\'e et al.(2019)]{dave19} Dav\'e R., Angl\'es-Alc\'azar D., Narayanan D., Li Q., Rafieferantsoa, M. H., Appleby S., 2019, MNRAS, 486, 2827

\bibitem[Dav\'e et al.(2020)]{dave20} Dav\'e, R., Crain, R.A., Stevens, A.R.H., Narayanan, D., Saintonge, A., Catinella, B. \& Cortese, L. 2020, MNRAS, 497, 146

\bibitem[Dekel et al.(2009)]{dekel} Dekel, A., Birnboim, Y., Engel, G., et al.
    2009, Natur, 457, 451

\bibitem[Dessauges-Zavadsky \& Adamo(2018)]{dessuages18} Dessauges-Zavadsky, M., Adamo, A, 2018, MNRAS, 479L, 118


\bibitem[Duc, Bournaud, \& Masset(2004)]{duc} Duc, P.-A., Bournaud, F., \& Masset, F. 2004, A\&A, 427, 803

\bibitem[Duncan et al.(2019)]{duncan19} Duncan, K., Conselice, C.J., Mundy, C. et al. 2019, ApJ, 876, 110


\bibitem[Elmegreen(2018)]{elm18} Elmegreen, B.G. 2018, ApJ, 869, 119

\bibitem[Elmegreen \& Efremov(1997)]{elmegreen97} Elmegreen, B.G., \& Efremov, Y.N. 1997, ApJ, 480, 235

\bibitem[Elmegreen \& Burkert(2010)]{elmburk} Elmegreen, B.G. \& Burkert, A. 2010, ApJ, 712, 294

\bibitem[Elmegreen \& Elmegreen(2005)]{elm05} Elmegreen, B.G. \& Elmegreen, D.M. 2005, ApJ, 627, 632

\bibitem[Elmegreen \& Elmegreen(2006b)]{elm06b} Elmegreen, B.G. \& Elmegreen, D.M. 2006, ApJ, 650, 644  %

\bibitem[Elmegreen et al.(2009b)]{elm09b} Elmegreen, B.G, Elmegreen, D.M., Fernandez, M.X., \& Lemonias, J.J. 2009, ApJ, 692, 12

\bibitem[Elmegreen et al.(1995)]{elm95} Elmegreen, B.G., Sundin, M., Kaufman, M., Brinks, E., \& Elmegreen, D.M. 1995, ApJ, 453, 139

\bibitem[Elmegreen \& Elmegreen(2006a)]{elm06} Elmegreen, D.M., \& Elmegreen, B.G., 2006, ApJ, 651, 676  
\bibitem[Elmegreen et al.(2007a)]{elm07a} Elmegreen, D.M., Elmegreen, B.G., Ferguson, T., \& Mullan, B. 2007, ApJ, 663, 734  

\bibitem[Elmegreen, Elmegreen, \& Hirst(2004)]{elm04} Elmegreen, D.M., Elmegreen, B.G., \& Hirst, A. 2004, ApJ, 604, L21

\bibitem[Elmegreen et al.(2007b)]{elm07b} Elmegreen, D.M., Elmegreen,  B.G., Ravindranath, S., \&  Coe, D. 2007, ApJ, 658, 763

\bibitem[Elmegreen et al.(2009a)]{elm09a} Elmegreen, D.M., Elmegreen, B.G., Marcus, M.T., Shahinyan, K., Yau. A.,\& Peterson, M. 2009, ApJ, 701, 306

\bibitem[Elmegreen \& Elmegreen(2014)]{elm14} Elmegreen, D.M. \& Elmegreen, B.G., 2014, ApJ,  781, 11

\bibitem[Evans et al.(2008)]{evans} Evans, A. S., Vavilkin, T., Pizagno, J., et al. 2008, ApJL, 675, L69

\bibitem[Fensch et al.(2017)]{fensch17} Fensch, J., Renaud, F., Bournaud, F., et al. 2017, MNRAS, 465, 1934


\bibitem[Ferreira et al.(2020)]{ferreira20} Ferreira, L., Conselice, C., Duncan, K., Cheng, T.-Y., Griffiths, A., \&  Whitney, A. 2020, ApJ, 895, 115

\bibitem[F\"{o}rster Schreiber et al.(2006)]{fs06} F\"{o}rster Schreiber, N., Genzel, R., Lehnert, N., Bouch\'{e}, N., Verma, A., et al. 2006, ApJ, 645, 1062

\bibitem[F\"{o}rster Schreiber et al.(2011)]{fs11} F\"{o}rster Schreiber, N., Shapley, A.E., Genzel, R., Bouch\'{e}, N., Cresci, G., et al. 2011, ApJ, 739, 45

\bibitem[Giavalisco et al.(2004)]{gia}Giavalisco, M., Ferguson, H., Koekemoer, A., et al. 2004,  ApJ, 600, L103

\bibitem[Grogin et al.(2011)]{grogin} Grogin, N., Kocevski, D., Faber, S., et al. 2011, ApJS, 197, 35

\bibitem[Guo et al.(2015)]{guo15} Guo, Y., Ferguson, H., Bell, E., Koo, D., Conselice, C., et al. 2015, ApJ, 800, 39

\bibitem[Guo et al.(2018)]{guo18} Guo, Y., Rafelski, M., Bell, E., Conselice, C., Dekel, A., et al. 2018, ApJ, 853, 108

\bibitem[Hopkins, Croton \& Bundy(2010)]{hopkins10} Hopkins, P.F., Croton, D., Bundy, K. 2010, ApJ, 724, 915

\bibitem[Hibbard \& Yun(1999)]{hibbard} Hibbard, J. E., \& Yun, M. S. 1999, AJ, 118, 162

\bibitem[Humphreys \& Davidson(1979)]{hump} Humpreys, R.M. \& Davidson, K. 1979, ApJ, 232, 409

\bibitem[Hung et al.(2016)]{hung16} Hung, C.-L., Hayward, C.C., Smith, H.A., Ashby, M.L.N., Lanz, L., Martínez-Galarza,
J.R., Sanders, D. B., Zezas, A. 2016, ApJ, 816, 99

\bibitem[Jiang, Wang, \& Qu(2011)]{jiang} Jiang, X., Wang, J., \& Gu, Q. 2011, MNRAS, 418, 1753

\bibitem[Kaufman et al.(1997)]{kauf95} Kaufman, M., Brinks, E., Elmegreen, D.M., Thomasson, M., Elmegreen, B.G., Struck, C., \&
Klar\'{i}c, M.  1997, AJ, 114, 2323

\bibitem[Kaufman et al.(2012)]{kauf12} Kaufman, M., Grupe, D., Elmegreen, B.G., Elmegreen, D.M., Struck, C., \& Brinks, E. 2012, AJ, 144, 156

\bibitem[Kere\v{s} et al.(2005)]{keres05} Kere\v{s}, D., Katz, N., Weinberg, D. H., \& Dav\'{e}, R. 2005, MNRAS, 363, 2

\bibitem[Kere\v{s} et al.(2009)]{keres09} Kere\v{s}, D., Katz, N., Fardal, M., Dav\'e,
    R., \& Weinberg, D. H. 2009, MNRAS, 395, 160

\bibitem[Knapen, Cisternas \& Querejeta(2015)]{knapen15} Knapen, J. H., Cisternas, M., \& Querejeta, M. 2015, MNRAS, 454, 1742

\bibitem[Koekemoer, Faber, Ferguson et al.(2011)]{koek} Koekemoer, A., Faber, S., Ferguson, H. et al. 2011, ApJS, 197, 36

\bibitem[K\"{o}nig et al.(2016)]{konig} K\"{o}nig, S., Aalto, S., Muller, S., Gallagher III, J. S.,  Beswick, R. J., Xu, C. K., \& Evans, A. 2016, A\&A, 594,A70

\bibitem[Krumholz \& McKee(2019)]{krumholz19} Krumholz, M.R. \& McKee, C.F. 2019, ARA\&A, 57, 227

\bibitem[Larson et al.(2020)]{larson} Larson, K.L., D\'{i}az-Santos, T., Armus, L., Privon, G.C., Linden, S.T. et al. 2020, ApJ, 888, 92

\bibitem[Lee et al.(2021)]{lee} Lee, J.C. et al. 2021, in preparation

\bibitem[Linden et al.(2017)]{lin} Linden, S.T., Evans, A.S., Rich, J., Larson, K.L., Armus, L. et al. 2017, ApJ, 843, 91

\bibitem[Lotz et al.(2008) ]{lotz08} Lotz, J. M., Jonsson, P., Cox, T. J., \& Primack, J. R. 2008, MNRAS, 391, 1137


\bibitem[Lotz, Primack, \& Madau(2004)]{lotz04} Lotz, J. M., Primack, J., \& Madau, P. 2004, AJ, 128, 163

\bibitem[Mantha et al.(2018)]{mantha18} Mantha, K.B., McIntosh, D.H., Brennan, R.,  et al. 2018, MNRAS, 475, 1549

\bibitem[Martin et al.(2017)]{martin17} Martin, G., Kaviraj, S., Devriendt, J. E. G., Dubois, Y., Laigle, C., Pichon, C. 2017, MNRAS, 472L, 50


\bibitem[Menacho et al.(2019)]{menacho} Menacho, V., Ostlin, G., Bik, A., Della Bruna, L., Melinder, J., Adamo, A., Hayes, M., Herenz, E.C., \& Bergvall, N. 2019, MNRAS, 487, 3183

\bibitem[Michiyama et al.(2018)]{mich} Michiyama, T., Iono, D., Sliwa, K., Bolatto, A., Nakanishi, K. et al. 2018, ApJ, 868, 95

\bibitem[Mihos(1995)]{mihos} Mihos, C. 1995, ApJ, 438, L75

\bibitem[Mirabel et al.(1992)]{mir} Mirabel, I. F., Dottori, H., \& Lutz, D. 1992, A\&A, 256, L19

\bibitem[Mok et al.(2019)]{mok19} Mok, A., Chandar, R., \& Fall, S.M. 2019, ApJ, 872, 93

\bibitem[Mok et al.(2020)]{mok} Mok, A., Chandar, R., \& Fall, S.M. 2020, ApJ, 893, 135

\bibitem[Mullan et al.(2011)]{mullan} Mullan, B., Konstantopoulos, I.S., Kepley, A.A., Lee, K.H., Charlton, J. et al. 2011, ApJ, 731, 93

\bibitem[Mundy et al.(2017)]{mundy17} Mundy, C.J., Conselice, C.J., Duncan, K.J., Almaini, O.,
Häussler, B., Hartley, W.G. 2017, MNRAS, 470, 3507


\bibitem[Ocvirk et al.(2008)]{ocvirk08} Ocvirk, P., Pichon, C., \& Teyssier, R. 2008, MNRAS, 390, 1326

\bibitem[O'Leary et al.(2020)]{oleary20} O’Leary, J.A., Moster, B.P. Naab, T., Somerville, R.S. 2020, arXiv:2001.02687v2

\bibitem[Pearson et al.(2019a)]{pearson19a} Pearson, W. J., Wang, L., Trayford, J. W., Petrillo, C. E., \& van der Tak, F. F. S. 2019, A\&A, 626, A49

\bibitem[Pearson et al.(2019b)]{pearson19b} Pearson, W. J., Wang, L., Alpaslan, M., et al. 2019, A\&A, 631A, 51 

\bibitem[Pillepich et al.(2018)]{pillepich18} Pillepich A. et al., 2018, MNRAS, 473, 4077

\bibitem[Price et al.(2020)]{price} Price, S.H., Kriek, M., Barro, G., Shapley, A.E., Reddy, N.A., et al. 2020, ApJ, 894, 91

\bibitem[Renaud et al.(2014)]{renaud14} Renaud, F.; Bournaud, F.; Kraljic, K.; Duc, P.-A. 2014, MNRAS, 442L, 33

\bibitem[Ribeiro et al.(2017)]{ribeiro17} Ribeiro, B., Le F\`evre, O., Cassata, P. et al. 2017, A\&A, 608A, 16
\bibitem[Rix et al.(2004)]{rix} Rix, H.-W., Barden, M., Beckwith, S., et al. 2004, ApJS, 152, 163

\bibitem[Sakamoto et al.(2014)]{saka} Sakamoto, K., Aalto, S., Combes, F., Evans, A., \& Peck, A. 2014, ApJ, 797, 90

\bibitem[S\'{a}nchez Almeida et al.(2014)]{sanchez} S\'{a}nchez Almeida, J., Elmegreen, B. G., Mu\~{n}oz-Tu\~{n}\'{o}n, C., \& Elmegreen, D. M. 2014, A\&ARv, 22, 71

\bibitem[Sancisi et al.(2008)]{sancisi08} Sancisi, R., Fraternali, F., Oosterloo, T., van der Hulst, T. 2008, A\&ARv, 15, 189

\bibitem[Sanders \& Tacconi(1996)]{sanders96} Sanders, D. B., \& Mirabel, I. F.  1996, ARA\&A, 34, 749

\bibitem[Schaye et al.(2010)]{schaye10} Schaye, J., Dalla Vecchia, C., Booth, C. M., et al. 2010, MNRAS, 402, 1536

\bibitem[Schaye et al.(2015)]{schaye15} Schaye, J., Crain, R. A., Bower, R. G., et al. 2015, MNRAS, 446, 521

\bibitem[Scoville et al.(1986)]{sco} Scoville, N. Z., Sanders, D. B.,  Sargent, A. I., Soifer, B. T.,  Scott, S. L.,  \& Lo,  K. Y.  1986, ApJ, 311, L47

\bibitem[Simons, Kassin, \& Weiner(2017)]{simons} Simons, R. C., Kassin, S. A., Weiner, B. J., et al. 2017, ApJ, 843, 46

\bibitem[Simons et al.(2019)]{simons19} Simons, R.C., Kassin, S.A., Snyder, G.F., et al. 2019, ApJ, 874, 59


\bibitem[Snyder et al.(2017)]{snyder17} Snyder, G.F., Lotz, J.M., Rodriguez-Gomez, V., Guimar\~aes, R., Torrey, P., Hernquist, L. 2017, MNRAS, 468, 207
\bibitem[Snyder et al.(2019)]{snyder19} Snyder,G.F., Rodriguez-Gomez,V., Lotz,J.M., etal. 2019, MNRAS, 486, 3702


\bibitem[Spergel et al.(2003)]{spergel} Spergel, D., Verde, L., Peiris, H.V., Komatsu, E., Nolta, M.R., Bennett, C.L., Halpern, M., Hinshaw, G., Jarosik, N., \& Kogut, A.  2003, ApJS, 148, 175

\bibitem[Struck(1999)]{struck} Struck, C. 1999, PhR, 321, 1

\bibitem[Tacconi et al.(2002)]{tacconi02} Tacconi, L. J., Genzel, R., Lutz, D., Rigopoulou, D., Baker, A. J., Iserlohe, C., Tecza, M. 2002, ApJ, 580, 73

\bibitem[Tacconi et al.(2010)]{tacc} Tacconi, L.J., Genzel, R., Neri, R., Cox, P., Cooper, M.C., et al. 2010, Nature, 463, 781

\bibitem[Tacconi et al.(2008)]{tacc08} Tacconi, L.J., Genzel, R., Smail, I., Neri, R., Chapman, S., et al. 2008, ApJ, 680, 246

\bibitem[Toomre \& Toomre(1972)]{toomre} Toomre, A., \& Toomre, J. 1972, ApJ, 178, 623

\bibitem[U et al.(2012)]{u2012} U, V., Sanders, D.B., Mazzarella, J.M., Evans, A.S., Howell, J.H., Surace, J.A. et al. 2012, ApJS, 203, 9

\bibitem[van den Bergh et al.(1996)]{vdb} van den Bergh, S., Abraham, R. G., Ellis, R. S., Tanvir, N. R., Santiago, B. X.,
\& Glazebrook, K. G. 1996, AJ, 112, 359

\bibitem[Ventou et al.(2019)]{ventou19} Ventou, E., Contini, T., Bouch\'e, N., et al.  2019, A\&A, 631A, 87


\bibitem[Verbeke et al.(2014)]{verb} Verbeke, R., De Rijcke, S., Koleva, M., et al. 2014, MNRAS, 442, 1830


\bibitem[Wen \& Zheng(2016)]{wen16} Wen, Z.Z. \& Zheng, X.Z. 2016, ApJ, 832, 90

\bibitem[Wheeler et al.(2020)]{wheeler} Wheeler, J., Glenn, J., Rangwala, N., \& Fyhrie, A. 2020, ApJ, 896, 43

\bibitem[Whitmore, Chandar, \& Fall(2007)]{whit07} Whitmore, B., Chandar, R., \& Fall, M. 2007, AJ, 133, 1067

\bibitem[Whitmore et al.(2010)]{whit} Whitmore, B., Chandar, R., Schweizer, F., Rothbery, B., Leitherer, C., Rieke, M., Rieke, G., Blair, W.P., Mengel, S., \& Alonso-Herrero, A. 2010, AJ, 140, 75

\bibitem[Whitmore et al.(2021)]{whit21} Whitmore, B., Rupali, C., Calzetti, D., Lee, J., Cook, D., Elmegreen D., Mok, A., Ubeda, L., \& White, R.,   in preparation

\bibitem[Whitmore et al.(1999)]{whit99} Whitmore, B., Zhang, Q., Leitherer, C., Fall, M., Schweizer, F., \& Miller, B.W. 1999, AJ, 118, 1551

\bibitem[Wilson et al.(2019)]{wilson19} Wilson, T.J., Shapley, A.E., Sanders, R.L., et al 2019, ApJ, 874 18

\bibitem[Wisnioski et al.(2019)]{wis19} Wisnioski, E., F\"{o}rster Schreiber, N. M., Fossati, M., et al. 2019, ApJ, 886, 124

\bibitem[Wisnioski et al.(2015)]{wis15}Wisnioski, E., F\"{o}rster Schreiber, N. M., Wuyts, S., et al. 2015, ApJ, 799, 209

\bibitem[Xu et al.(2020)]{xu20} Xu, K., Liu, C., Jing, Y., Wang, Y., \& Lu, S. 2020, ApJ, 895, 100

\bibitem[Zanella et al.(2019)]{zanella19} Zanella, A., Le Floc'h, E., Harrison, C. M., et al. 2019, MNRAS, 489, 2792

\bibitem[Zaragoza-Cardiel et al.(2018)]{zara} Zaragoza-Cardiel, J., Smith, B.J., Rosado, M., Beckman, J.E., Bitsakis, T., Camps-Fari\~{n}a, A., Font, J., \& Cox, I.S. 2018, ApJS, 234, 35

\bibitem[Zepf et al.(1999)]{zepf} Zepf, S.E., Ashman, K.M., English, J., Freeman, K.C., \& Sharples, R.M. 1999, AJ, 118, 752



\end{thebibliography}
\end{document}